\documentclass[review]{elsarticle}

\usepackage{lineno,hyperref}
\modulolinenumbers[5]

\usepackage{amsmath,physics} 
\usepackage{amsfonts}
\usepackage{textcomp} 
\usepackage{booktabs,tabularx}
\usepackage{graphicx}
\usepackage{xcolor}
\usepackage{subfigure}
\usepackage{lscape}
\usepackage{rotating} 
\usepackage{mathtools}
\DeclarePairedDelimiter\ceil{\lceil}{\rceil} 

\DeclareMathAlphabet{\mathbbmsl}{U}{bbm}{m}{sl}

\begin{document}

\begin{frontmatter}


\title{Multi-stage Power Scheduling Framework for Data Center with Chilled Water Storage in Energy and Regulation Markets}


\author[mymainaddress]{Yangyang Fu}

\author[mymainaddress]{Xu Han}

\author[mymainaddress]{Jessica Stershic}

\author[mymainaddress,mythirdaddress]{Wangda Zuo\corref{mycorrespondingauthor}}
\cortext[mycorrespondingauthor]{Corresponding author}
\ead{Wangda.Zuo@Colorado.edu}

\author[mymainaddress,mythirdaddress]{Kyri Baker}
\author[myfourthaddress]{Jianming Lian}


\address[mymainaddress]{Department of Civil, Environmental and Architectural Engineering, University of Colorado, Boulder, USA}
\address[mythirdaddress]{National Renewable Energy Laboratory, Golden, USA}
\address[myfourthaddress]{Pacific Northwest National Laboratory, Richland, USA}

\begin{abstract}
Leveraging electrochemical and thermal energy storage systems has been proposed as a strategy to reduce peak power in data centers.
Thermal energy storage systems, such as chilled water tanks, have gained increasing attention in data centers for load shifting due to their relatively small capital and operational costs compared to electrochemical energy storage. 
However, there are few studies investigating the possibility of utilizing thermal energy storage system with resources to provide ancillary services (e.g., frequency regulation) to the grid. 
This paper proposes a synergistic control strategy for the data center with a chilled water storage providing frequency regulation service by adjusting the chiller capacity, storage charging rate, and IT server CPU frequency. 
Then, a three-stage multi-market scheduling framework based on a model predictive control scheme is developed to minimize operational costs of data centers participating in both energy and regulation markets.
The framework solves a power baseline scheduling problem, a regulation reserve problem, and a real-time power signal tracking problem sequentially.
Simulation results show that utilizing the thermal energy storage can increase the regulation capacity bid, reduce energy costs and demand charges, and also harvest frequency regulation revenues. 
The proposed multi-market scheduling framework in a span of two days can reduce the operational costs up to 8.8\% (\$1,606.4) compared to the baseline with 0.2\% (\$38.7) energy cost reduction, 6.5\% (\$1,179.4) from demand reduction, and 2.1\% (\$338.3) from regulation revenues.

\end{abstract}


\end{frontmatter}


\section{Introduction and Motivation}\label{sec:chap6-introduction}
Energy storage systems have been proposed to reduce the peak power in data centers. 
The energy storage system can be charged during a period of low power use and discharged during peak power use. This allows data centers to provide demand response for the power grid while meeting the electrical demand of the data center.
Two types of energy storage systems are generally used in data centers: electrochemical energy storage system (EESS) and thermal energy storage system (TESS).
The EESS, such as uninterruptible power supply (UPS) batteries, have been commonly used in data centers. 
Many research have focused on shaving the data center power demand using EESS.
However, these batteries have several limitations. 
First, batteries in data centers have limited capacities. 
The design purpose of batteries in data centers is to serve critical equipment (e.g., IT equipment, air fans etc.) for a short period during an emergency situation, such as a power outage to allow backup generators to start-up. 
They are typically sized to provide power for 5-15 minutes with a small storage capacity~\cite{fu2019equation-b} .
Second, the performance of conventional batteries in data centers deteriorates quickly when frequent charging and discharging is required.
Third, batteries are currently expensive, requiring high capital and recycling costs.
Therefore, many researchers have shifted their attention to find an alternative for the EESS.

TESS, especially chilled water storage, because of its relatively low costs, has gained attention for power shaving in data centers. 
Many industrial data centers, such as a Google data center in Taiwan, have adopted TESS to avoid high operational costs during on-peak periods~\cite{dc2019google}.  
There is also research focusing on utilizing TESS for data center power management.
Zheng et al.~\cite{zheng2014exploiting} proposed a strategy of using TESS to shave the data center power profile, which significantly reduces the capital and operational costs.
Zhang et al.~\cite{zhang2012testore} designed and evaluated a cooling strategy that exploits EESS and TESS techniques to cut the electricity bill for data center cooling, without causing servers in the data center to overheat.
Oro et al. \cite{oro2015overview} reviewed the energy saving potentials of direct air cooling and TESS in data centers and concluded that when using TESS in combination with an off-peak electricity tariff, the operational cooling cost can be drastically reduced.

However, no studies have considered the potential of using TESS in data centers to provide ancillary services such as frequency regulation (FR) to the power grid.
FR requires fast response by demand side resources (DSRs) to the power grid signal.
A typical fast-responding resource in data centers is the server, and a lot of research has been proposed to manipulate servers to provide fast demand response service and harvest the benefits \cite{wang2019frequency,li2013data}.
Nevertheless, they barely considered the cooling system, let alone thermal storage.
There are multiple reasons, one of which may be concerns over the large thermal time constant (e.g, hours) of the cooling system and TESS compared with the responding time scale (e.g., seconds) required by FR.

Data centers themselves reveal numerous opportunities that can overcome those concerns.
First, data centers can operate under a broad range of temperatures, which will result in a large range of power load. 
For example, American Society of Heating, Refrigerating, Air-Conditioning Engineers (ASHRAE) categorizes data centers into four types (A1-A4) based on their requirements of the thermal environment. 
A Class A1 data center typically provides mission critical operations and requires a tightly controlled thermal environment.
ASHRAE suggests that the allowable supply air temperature in a Class A1 data centers should be within the range of 15 \textdegree C to 32 \textdegree C \cite{ashrae2015thermal}. 
Second, a data center is an excellent DR candidate consisting of slow-responding resources (e.g., cooling system) and fast-responding resources (e.g., servers).
Fu et al.~\cite{fu2019assessments} show that when providing fast demand response the delays caused by slow-responding resources can be compensated by fast-responding resources if any.
Therefore, utilizing TESS and servers simultaneously might be capable of providing FR service. 

\begin{figure}[htbp]
\centering
\subfigure[Extra demand imposed to FR baseline]{%
\resizebox*{6cm}{!}{\includegraphics{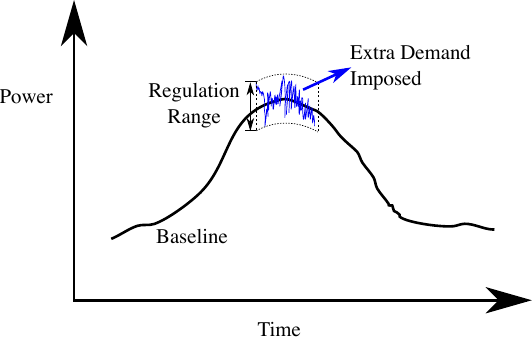}}}
\subfigure[Normalized extra demand of 30 minutes during July, 2018 in PJM]{%
\resizebox*{6cm}{!}{\includegraphics{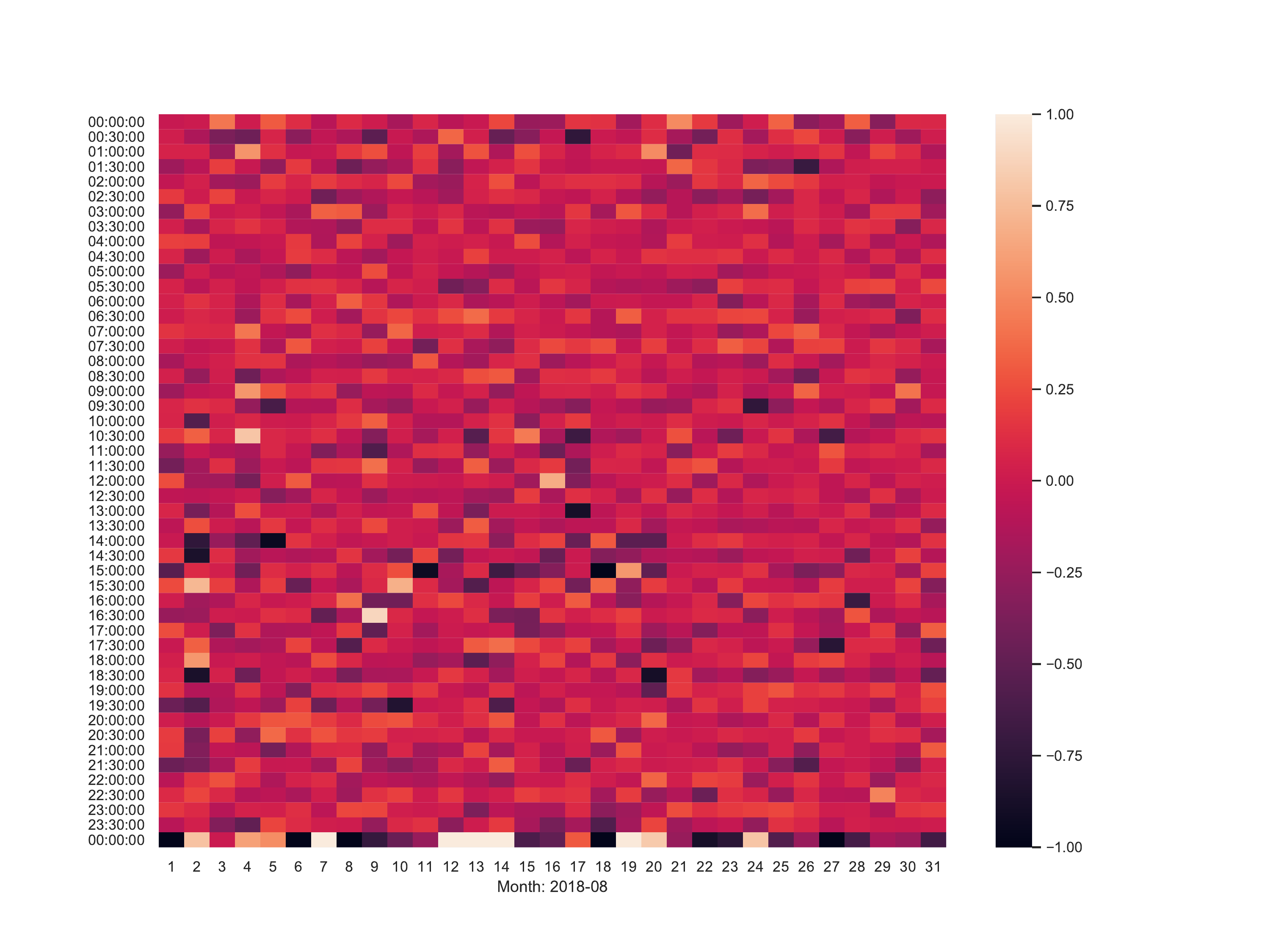}}}
\caption{Extra demand imposed by FR signal} 
\label{fig:fr-extra-demand}
\end{figure}

Despite the above-mentioned unique opportunities, data centers face some other challenges as well. 
One of them is that the data center room has large internal heat gains discharged by servers. 
The thermal mass in the room such as walls and racks can be quickly charged or discharged, which leads to negligible thermal inertia in the room.  
A change in the cooling system or the IT equipment can be quickly reflected in the data center room.
Therefore, if the cooling and IT system are to provide grid services, they need to be carefully designed to avoid violations of desired thermal conditions.
Second, data centers have a large power demand due to its peak workload. 
The large power demand can provide a large FR capacity~\cite{fu2019assessments}, but the large capacity may introduce extra power demand when providing FR at the peak workload as illustrated in Figure~\ref{fig:fr-extra-demand}(a). 
The extra power demand imposed on the peak depends on the FR signal.
Figure~\ref{fig:fr-extra-demand}(b) shows a map of the normalized extra demand of 30 minutes when providing FR during July, 2018 in PJM.
Although the FR signal over a long time horizon (e.g., hours) sums to zero, the sum can be positive at a smaller time scale such as 5 or 30 minutes, which is a typical time slot for demand charges by utilities.
These positive sums would lead to additional demand costs when imposed on the peak~\cite{fu2020multi}. 
The additional demand costs might even be larger than the FR revenues, which compromises the final benefits of providing demand response.

This paper proposes a new control framework that enables TESS in data centers to provide FR service and a multi-market optimization framework to harvest benefits from both energy and regulation markets without introducing extra demand charges. 
First, essential background about a typical chilled-water TESS and FR service are introduced in Section~\ref{sec:chap6-background}.
Then, a three-stage multi-market scheduling framework is proposed to enable data centers to minimize their costs in the energy and regulation markets.
The three-stage framework sequentially solves a power baseline scheduling problem, a regulation reserve problem, and a real-time power signal tracking problem at different time scales.
The power baseline scheduling and regulation reserve problem is formulated and solved using a model predictive control (MPC) scheme and the real-time power signal tracking is realized by using a new closed-loop synergistic control strategy.
The new synergistic control strategy enables FR service by adjusting the chiller capacity, storage charging rate, and IT server CPU frequency simultaneously.
Section~\ref{sec:chap6-case-study} then numerically demonstrates the control performance of the proposed three-stage multi-market optimization framework.
Finally, the paper is concluded in Section~\ref{sec:chap6-conclusions}.

\textit{Notation}: Throughout the article, $a^t$ represents the value of a variable $a$ at time $t$,$\widetilde{a}$ represents the predicted value of variable $a$, bold letters are used to denote sequences over time, e.g.,$\boldsymbol{p} ^{t}= [p^{t}, p^{t+1}, . . ., p^{t+n}]$, and the capital Greek symbols such as $\Psi$ denote implicit functions.

\section{Background}\label{sec:chap6-background}
\subsection{Chilled-water Thermal Energy Storage System}\label{subsec:chap6-tess-system}
\subsubsection{Storage Charging and Discharging}
\begin{figure}[htbp]
\centering
\subfigure[Charging]{%
\resizebox*{6cm}{!}{\includegraphics{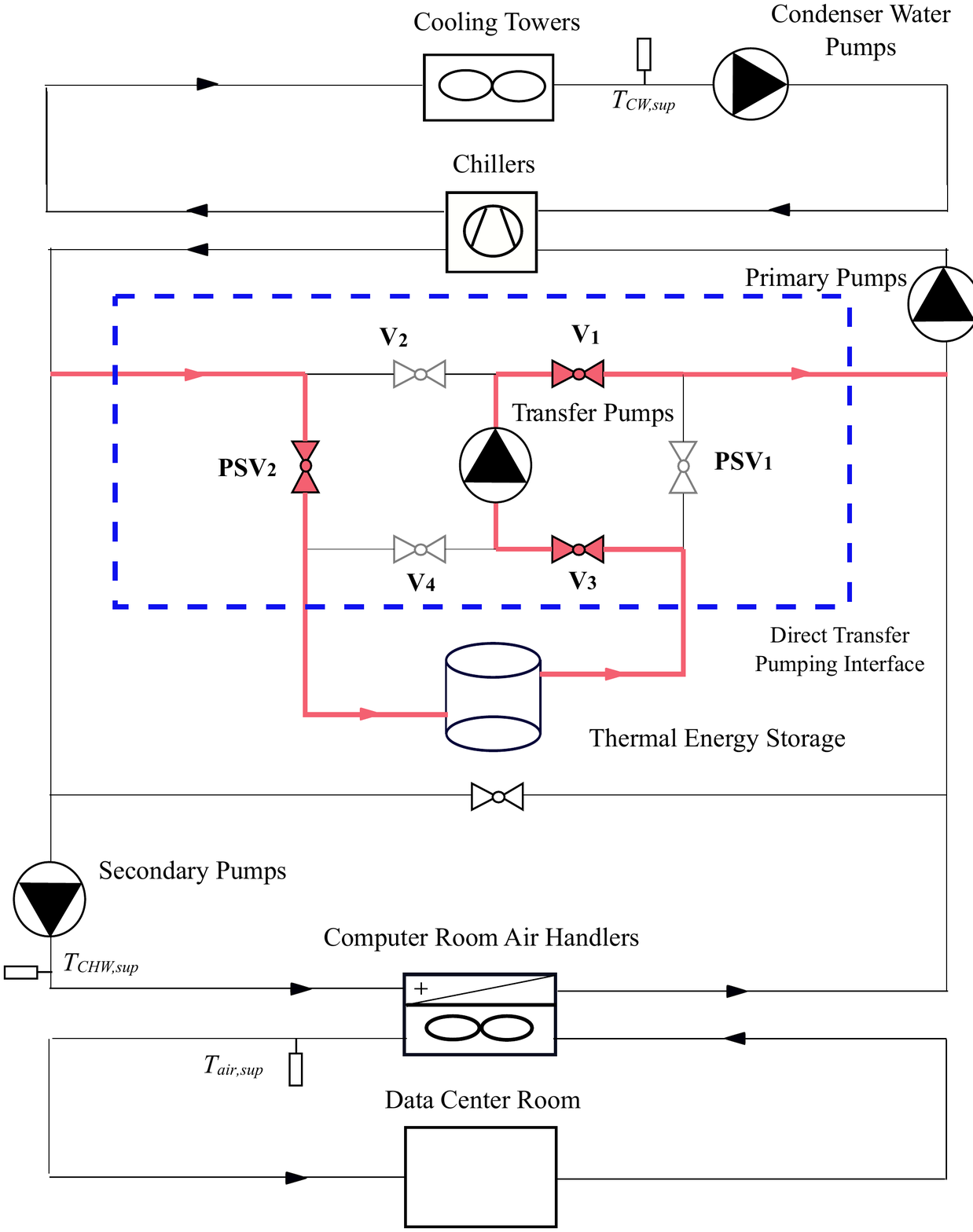}}}
\subfigure[Discharging]{%
\resizebox*{6cm}{!}{\includegraphics{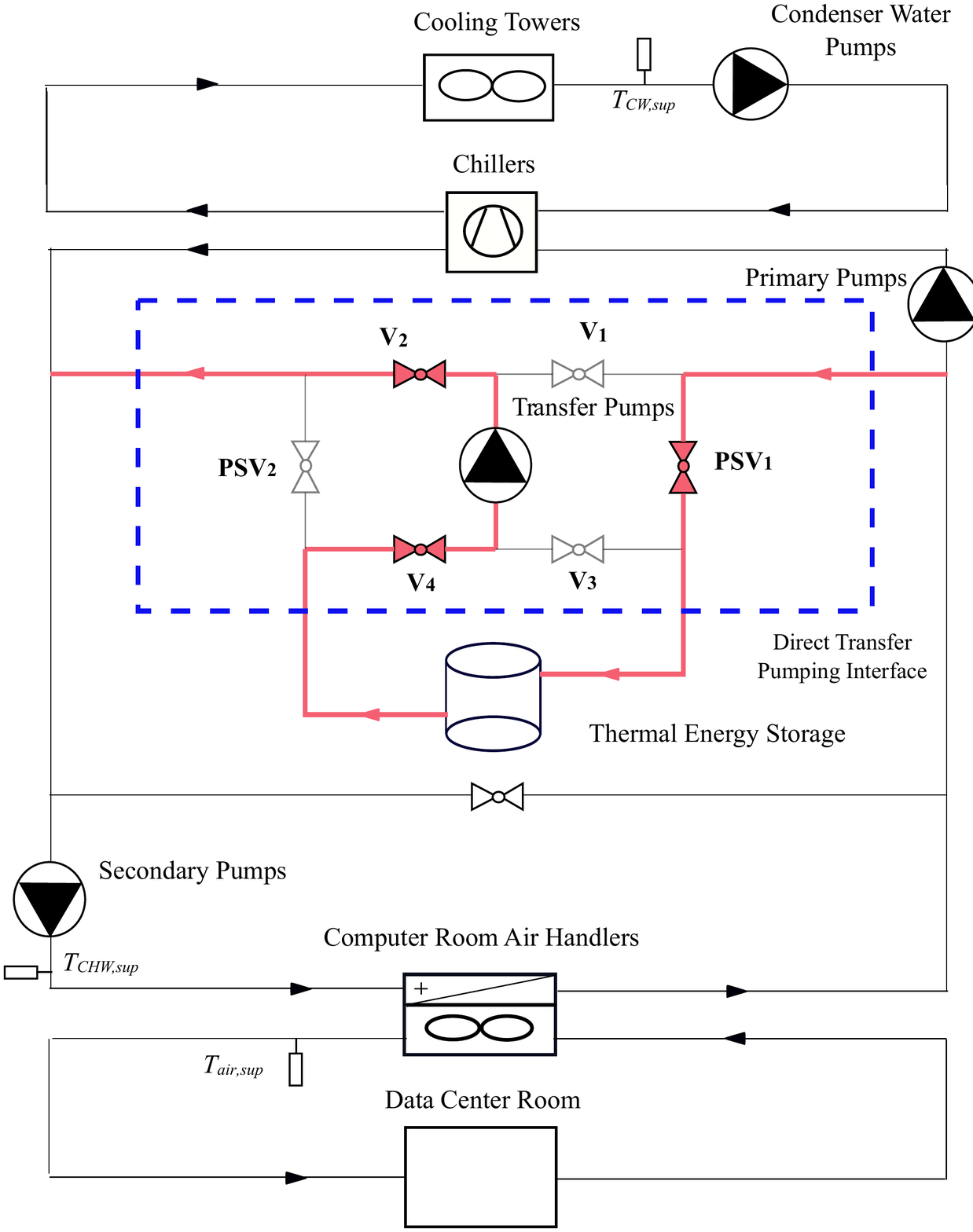}}}
\caption{Chilled water TESS serving a primary-secondary chilled water system} 
\label{fig:chap6-charging-discharging}
\end{figure}

One commonly-used cool storage system is known as a chilled water tank, which has been successfully applied in a variety of settings such as campus and district cooling systems, power generation, and emergency cooling systems~\cite{handbook2016hvac}.
A typical inside-the-plant connection scheme for a data center chilled water tank serving a primary/secondary chilled water system is shown in Figure~\ref{fig:chap6-charging-discharging}(a)~\cite{handbook2016hvac}.
The primary pumps operate at a constant speed, while the secondary pumps operate at various speeds.
Besides the traditional configuration of a primary-secondary pump system, a direct transfer pumping interface is installed to pump chilled water between the tanks and the chilled water system~\cite{bahnfleth1999analysis}.
The charging and discharging of a chilled water tank with vertical temperature stratification is realized by operating the direct transfer pump interface as shown in Table~\ref{tab:chap6-operate-charging} and Figure~\ref{fig:chap6-charging-discharging}.
The charging and discharging process can be characterized by the charging rate $u_s$ as defined in Section~\ref{subsubsec:chap6-tank-model}. 
The charging rate describes the "cooling energy" that is being added or removed from the storage each second.
A positive value means charging and a negative value means discharging.

\begin{table}[htbp]
\centering
\caption{Operations of charging and discharging storage}
\label{tab:chap6-operate-charging}
\begin{tabular}{@{}llllllll@{}}
\toprule
Actions     & V\textsubscript{1} & V\textsubscript{2} & V\textsubscript{3} & V\textsubscript{4} & PSV\textsubscript{1}    & PSV\textsubscript{2}    & Pumps     \\ \midrule
Charging    & open   & closed  & open   & closed  & closed & modulated & modulated \\
Discharging & closed  & open   & closed  & open   & modulated & closed & modulated \\
Idle        & open  & closed  & closed  & closed  & closed & closed & closed       \\ \bottomrule
\end{tabular}%
\end{table}

\subsection{Electricity Markets}
The frequency of electrical grids must be maintained at their nominal values (e.g., 60 Hz in United States) through the continuous real-time balancing of power demand and supply. 
The balancing in deregulated electric grids can be managed by regional transmission organizations (RTOs) or independent system operators (ISOs) via electrical markets. 
Towards this service as well as longer-term energy balancing, there are typically three types of electrical markets: capacity markets, energy markets, and ancillary services markets. 
These markets manage the power grid over time scales ranging from several years to seconds. For example, the capacity market looks at least one year ahead to ensure sufficient generation capacity online, while the energy market and ancillary service market focus on shorter time scales of hours to seconds based on the different services required.

The energy market usually procures day-ahead and hourly services to ensure that sufficient generation capacity is available on a day-ahead to one-hour-ahead basis.
Therefore, it usually consists of two markets: a day-ahead market and a real-time market.
The day-ahead market is used to determine which generators are scheduled to operate during each hour of the following day and at what level of output based on a projection of electricity demand in the following day. 

The ancillary service market allows the RTO/ISO to maintain the system frequency on a sub-hour basis.
There are many different types of ancillary services, including spinning/non-spinning reserve and regulation reserve etc. 
Note different countries and RTOs/ISOs have different terminologies about these services~\cite{rebours2007survey}.
Regulation reserve, also known as FR, represents capacity that can adjust its level of output within a few seconds in response to the fluctuations in the system frequency.
The fluctuations are typically caused by slow response, inaccurate automatic generation control, and forced outages of power plants etc.

\subsection{Frequency Regulation}\label{subsec:chap6-fr}
FR can be offered by FR resources such as generators on the supply side (which has traditionally been the case) or more recently, by DSRs on the demand side.
Providing FR means FR resources are willing to increase or decrease their output (generation for generators, and consumption for DSRs) by following a control signal generated by the market operator.

Different market operators adopt different FR policies. This study uses Pennsylvania-New Jersey-Maryland independent system operator territory, known as the PJM ISO. The remaining section introduces a few important and relevant features using PJM as an example. Details of PJM FR service can be found in \cite{llc2019pjm}. 
PJM divides FR resources into two categories: ramp-limited and capacity-limited. 
Ramp-limited resources respond slowly to FR signals, but with a large capacity. One example could be a coal-fired steam power plant.
Capacity-limited resources, including batteries, flywheels, and responsive loads, have small capacities, but can respond to FR signals quickly. 
PJM has developed two types of FR signals for these two resources: traditional regulation A signal (RegA) for ramp-limited resources and dynamic regulation D signal (RegD) for capacity-limited resources. 
Under these two FR signals, ramp-limited resources mostly get paid for their capacity and capacity-limited resources mostly get paid for their performance, which is defined in the following paragraph.

In the PJM regulation market, FR resources are required to maintain a minimum performance score of 0.4 in order to participate. 
The performance score $s$ is calculated as a composite score of accuracy, delay, and precision, which are shown below \cite{llc2019pjm}.
\begin{align}
    c_{sig, res} = \frac{COV(reg, res)}{\sigma_{reg} \sigma_{res}} \\
    s_{acc} = \max_{\delta = 0 - 5 ~\textrm{min}}(c_{reg,res(\delta)}) \\
    s_{del} = \abs{\frac{5 ~\textrm{min}-\delta^*}{5 ~\textrm{min}}} \label{eq:delay-score}\\
    s_{pre} = 1 - 
    \frac{1}{n}\sum\abs{\frac{res-reg}{\overline{reg}}} \\
    s = \frac{S_{acc}+S_{del}+S_{pre}}{3} \label{equ:pjm-performance-score}
\end{align}

In the above equations, $reg$ represents the regulation signal the DSRs receive from the electrical markets and $res$ represents the response signal the DSRs generate after control actions.
$c$, $COV$ and $\sigma$ are the correlation coefficient, covariance, and standard deviation of these two signals respectively.
In PJM, the response signal $res$ is recalculated with a time shift $\delta$ ranging from 0 to 5 minutes in an increment of 10 seconds, which leads to 31 response signals $res(\delta)$.
The accuracy score $s_{acc}$ is the maximum correlation coefficient $c$ between $reg$ and $res(\delta)$.
The delay score $s_{del}$ is calculated based on the delay time $\delta^*$ when the maximum accuracy score is obtained using Eq.~(\ref{eq:delay-score}). 
The precision score $s_{pre}$ is defined as the relative difference between the regulation and response signals, where $n$ is the number of samples in the hour and $\overline{reg}$ is the hourly average regulation signal.
The hourly final performance score $s$ is calculated as the weighted average of the three individual scores.

\section{Multi-market Scheduling Framework}\label{sec:multi-market-control}
\begin{figure}[h!]
   \centering
    \includegraphics[scale=0.2]{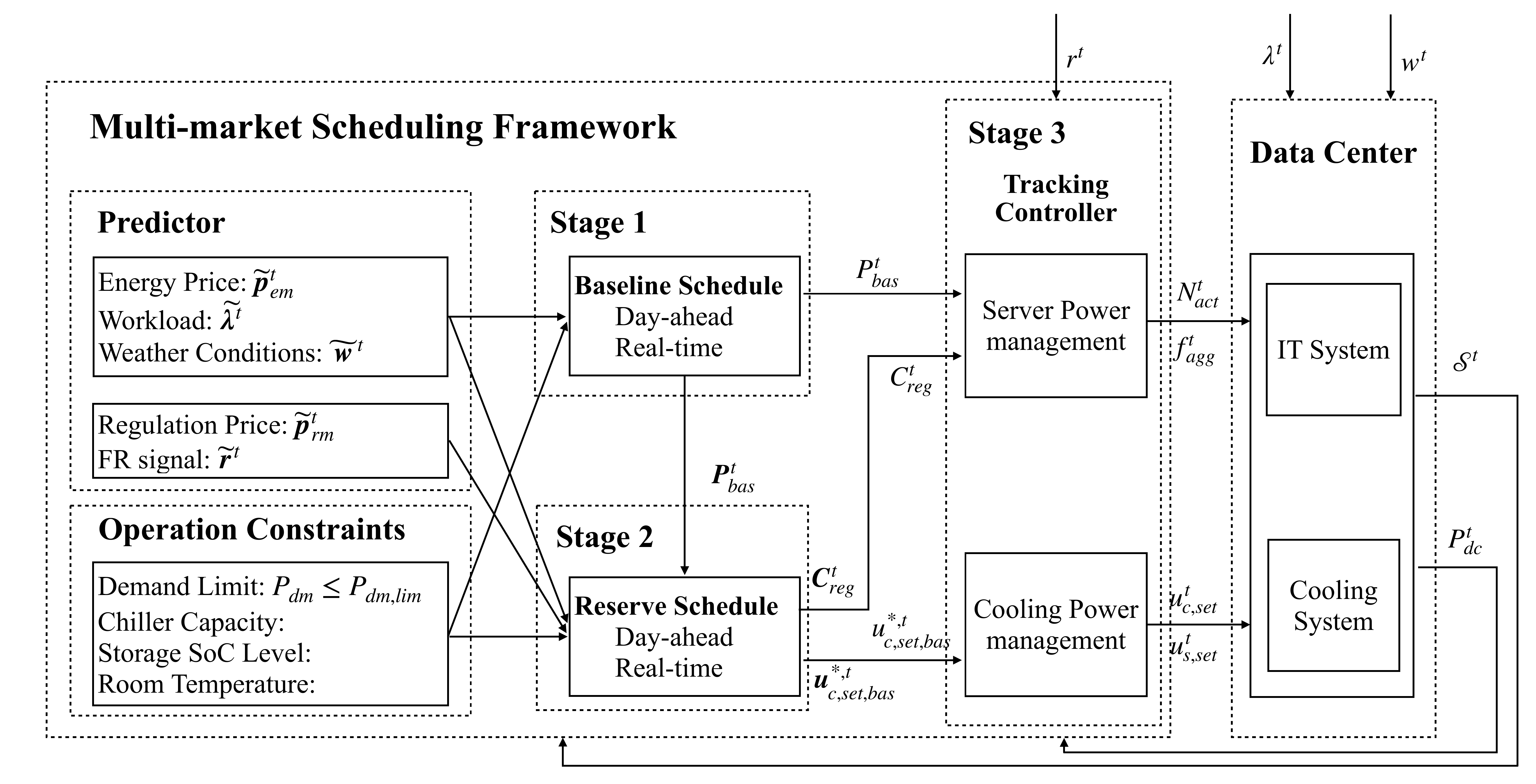}
    \caption{Diagram of multi-market optimization framework}
    \label{fig:chap6-mpc-framework}
\end{figure}

In this section, we propose a three-stage multi-market scheduling framework to enable data centers participation in the energy market and regulation market while minimizing their costs and maintaining the desired room thermal environment. 
In this framework, we consider the scheduling problem of energy baseline purchase and regulation reserve, and how to provide regulation services in data centers.
The energy and regulation markets are not modeled, therefore, we assume in this paper that the data center is a ``price taker".
The framework is depicted in Figure~\ref{fig:chap6-mpc-framework}, which consists of a predictor, a block for operation constraints, and a three-stage controller. 

The predictor provides predictions of system inputs at each time step $t$, including energy prices $\widetilde{\boldsymbol{p}}^{t}_{em}$, regulation reserve prices $\widetilde{\boldsymbol{p}}^{t}_{rm}$, data center workload $\widetilde{\boldsymbol{\lambda}}^{t}$, FR signal from the electric grid $\widetilde{\boldsymbol{r}}^{t}$, and the outdoor weather conditions $\widetilde{\boldsymbol{w}}^{t}$.
The block for operation constraints provides a set of system operation constraints such as the demand limit, chiller operation constraints, storage operation constraints, and required room thermal environments etc.

The \textit{Stage 1} controller decides on the baseline purchased on the day-head and hour-ahead energy market $\boldsymbol{P}^{t}_{bas}$ for time step $t$ of the following day.
Subsequently, it can then adjust the power baseline up to 1 hour in advance on the real-time market by placing a bid $P^{t|t-\delta}_{bas}$ at time $t-\delta$ for time $t$ based on new predictions.
For simplicity, here we assume the data center only participates in the real-time market and submit bids at time $t-1$ for time $t$.
The purchased power baseline $\boldsymbol{P}^{t}_{bas}$ over a horizon is then passed to the \textit{Stage 2} controller.
Details about power baseline scheduling are introduced in Section~\ref{subsec:stage1}.

The \textit{Stage 2} controller then schedules the reserved regulation capacity $\boldsymbol{C}^{t}_{reg}$ over a horizon for providing FR service.
Similar to the energy market, here we simplify the bidding process by only bidding into the real-time market, which means we only calculate and submit the bid one-hour ahead.
The regulation capacity comes from by regulating server CPU frequency and chiller capacity.
Details are included in Section~\ref{subsec:stage2}.

The calculated schedules $P^{t}_{bas}$, $C^{t}_{reg}$, and optimal $u^{t}_{c,set,bas}$ for time $t$ are then passed to the \textit{Stage 3} controller, which assures that the data center total measurement power $P_{dc}$ can track the power reference signal at the seconds-level.
The tracking is provided by adjusting the servers (number of activated servers $N^{t}_{act}$ and the aggregated frequency $f^{t}_{agg}$), chiller capacity $u^{t}_{c,set}$, and storage charging rate $u^{t}_{s,set}$. 
The system states $\mathcal{S}$ after adjusting those control inputs are then updated, and fed back to the multi-market scheduling framework for next time step. 
Detailed design of the tracking controller is provided in Section~\ref{subsec:stage3}.

\subsection{Stage 1: Baseline Schedule}\label{subsec:stage1}
The \emph{Baseline Schedule} calculates a power baseline profile for the next few hours, which is the power usage by the data center when not providing FR service.
The power baseline is then used by the tracking controller to calculate the reference power signal $P_{ref}$.
There are many rule-based strategies that can be utilized to generate the baseline.
One typical control strategy for TESS is the storage-priority control as detailed in~\cite{henze1997development}, which prioritizes the use of storage for discharging during on-peak periods and charging during off-peak periods. 
During off-peak periods, the chillers are operated at full capacity until the tank is either full or charged to the level that which the cumulative cooling load during the next on-peak period can be met without operating the chillers.
During on-peak periods, the chillers operate at a constant capacity for the entire on-peak period so that the storage is just depleted at the end of the on-peak period.
However, those rule-based control strategies might not be able to provide optimal control of storage to minimize operational costs.
This \textit{Stage 1} controller utilizes an MPC scheme to find the most economic power baseline profile with a power demand limit $P_{dm,lim}$ at each time step.
The MPC controller optimizes the control signal chiller capacity $\boldsymbol{u}^{t}_{c,set}$ over a finite prediction horizon $h$ at time $t-1$, and implements the optimal signal for time $t$. 
The operational constraints described above are considered here.
The MPC formulation for time $t$ is shown as follows.

\begin{eqnarray}
\textbf{Problem 1} \textit{:Baseline Schedule} & \nonumber\\
    \min_{\boldsymbol{u}^{t}_{c,set}} \quad & J^{t}(\boldsymbol{u}^{t}_{c,set}, \boldsymbol{u}^{t}_{s,set},\widetilde{\boldsymbol{p}}^{t}_{em},\widetilde{\boldsymbol{\lambda}}^t,\widetilde{\boldsymbol{w}}^t)\nonumber \\
    &  = E^{t}_{cos} + D^{t}_{pen} \label{equ:mpc1-obj}\\ 
    s.t. \quad &  \nonumber\\
\text{Building Constraints} & \boldsymbol{P}^{t}_{dc}, \mathcal{S}^{t} = \Gamma (\boldsymbol{u}^{t}_{c,set},\boldsymbol{u}^{t}_{s,set},\widetilde{\boldsymbol{\lambda}}^t,\widetilde{\boldsymbol{w}}^t) \label{equ:bc-system},\\ 
    &  \mathcal{S}_{min} \le \mathcal{S}^{t} \le \mathcal{S}_{max} \label{equ:bc-states}, \\   
\text{Control Constraints}  & u_{c,min} \leq \boldsymbol{u}^{t}_{c,set} \leq u_{c,max} \label{equ:cc-chiller-range}, \\
    & \boldsymbol{u}^{t}_{s,set} =  \widetilde{\boldsymbol{\dot{q}}}^t - \boldsymbol{u}^{t}_{c,set} \label{equ:cc-storage-set}, \\
    & u_{s,min} \leq \boldsymbol{u}^{t}_{s,set} \leq u_{s,max} \label{equ:cc-storage-range}
\end{eqnarray}

The cost function $J$ in this problem is defined as the total operational costs in terms of energy and demand, which is related to the control signals such as the chiller capacity setpoint, storage charging rate, and the predicted inputs such as energy price, workload, and outdoor weather conditions.
The cost has two parts: energy cost $E_{cos}$ and demand penalty $D_{pen}$.
The energy cost is calculated by Eq.~({\ref{equ:chap6-energy-cost}}). 
\begin{equation}\label{equ:chap6-energy-cost}
        E^{t}_{cos} = \widetilde{\boldsymbol{p}}^{t}_{em}\boldsymbol{P}^{t}_{dc}=\sum^{t+h}_{i=t} \widetilde{p}^i_{em}P^i_{dc}
\end{equation}
where $\widetilde{p}_{em}$ is the predicted energy price and $P_{dc}$ is the total power consumption for the data center. 
The calculation period starts from time $t$ and ends at $t+h$. 
The electric demand during the current prediction horizon is penalized by the demand price $p_{dm}$:
\begin{equation}\label{equ:chap6-demand-cost}
    D^t_{pen}  =  \max_{t,t+h}((\boldsymbol{P}_{dm}-P_{dm,lim}), 0)\cdot p_{dm},
\end{equation}
where $p_{dm}$ is the demand price, which is a fixed price. 
The $\boldsymbol{P}_{dm}$ is the predicted average power demand over each 30-minute interval, and $P_{dm,lim}$ is a user input, denoting the limit of required power demand.
If the demand exceeds a predefined demand limit, then the optimization cost function is penalized by the demand cost differences. 
Otherwise, no penalty is applied.
Note that $p_{dm}$ and $\boldsymbol{P}_{dm}$ are both utility specific and may vary from this definition for other energy markets. 

The constraints are composed of building constraints and control constraints. 
The building constraints assure that the building states are within a feasible boundary.
The building system dynamics are modelled as $\Gamma$, which can predict the total power consumption $\boldsymbol{P}_{dc}$ and system states $\mathcal{S}$.
The detailed description of $\Gamma$ is presented in Section~\ref{sec:dc-model}.
The output states include the room temperature $\boldsymbol{T}_{roo}$ and state-of-charge of the chilled water tank $\boldsymbol{SoC}$.
The states are bounded by their minimum and maximum allowable values, which can mathematically represented as Eq.~(\ref{equ:bc-states}).
The control constraints define the physical relationship and the allowable ranges of the control signals.
Eq.~(\ref{equ:cc-chiller-range}) describes the physical limits of the chiller capacity setpoint within a minimum of $u_{c,min}$ and a maximum of $u_{c,max}$, and Eq.~(\ref{equ:cc-storage-set}) and Eq.~(\ref{equ:cc-storage-range}) describe the charging limits of the storage tank. 
$u_{s,min}$ and $u_{s,max}$ are the maximum discharging and charging rates of the tank, which are physically constrained by the system design and operation, such as transfer pumps and temperatures in the tank.
$\widetilde{\boldsymbol{\dot q}}^t$ is the predicted cooling load in the data center.
Here, an assumption that servers disperse all the input power as thermal heat (cooling load) in the room is made using Eq.~\ref{equ:cool-load}.
The power baseline $\boldsymbol{P}^t_{bas}$ scheduled from this controller is then obtained from the following equation after solving the above optimization problem:
\begin{equation}\label{equ:base-power} 
    \boldsymbol{P}^t_{bas} = \boldsymbol{P}^{t}_{dc} 
\end{equation}

\subsection{Stage 2: Reserve Schedule}\label{subsec:stage2}
After the power baseline from the energy market is scheduled, the \textit{Stage 2} controller \textit{Reserve Schedule} decides on the reserve schedules for providing regulation service by maximizing the benefits from both the energy market and regulation market. 
This controller is also formulated using MPC schemes, calculating a sequence of schedules of the FR capacity bid $\boldsymbol{C}^{t}_{reg}$ and base chiller capacity setpoint $\boldsymbol{u}^{t}_{c,set,bas}$ for each time step $t$. 
But only the control signals $C^t_{reg}$ and $u^{t}_{u,set,bas}$ are passed to the \textit{Stage 3} controller.

The MPC cost function is formulated as follows.
For each time step $t$, the optimal base chiller capacity setpoint $\boldsymbol{u}^{*,t}_{c,set,bas}$ can be calculated from the following problem.

\begin{eqnarray}\label{equ:chap6-mpc-form}
\textbf{Problem 2} \textit{:Reserve Schedule} & \nonumber\\
    \min_{\boldsymbol{u}^{t}_{c,set,bas}} \quad & J^{t}(\boldsymbol{u}^{t}_{c,set,bas}, \widetilde{\boldsymbol{p}}^{t}_{em},\widetilde{\boldsymbol{p}}^{t}_{rm},\widetilde{\boldsymbol{\lambda}}^t,\widetilde{\boldsymbol{w}}^t,\boldsymbol{P}^t_{bas},\widetilde{\boldsymbol{r}}^t)\nonumber \\
    &  = E^{t}_{cos} + D^{t}_{pen} - R^{t}_{rev} \label{equ:mpc2-obj}\\ 
    s.t. \quad &  \nonumber\\
\text{Building Constraints} &  \text{Eq.~(\ref{equ:bc-system}), Eq.~(\ref{equ:bc-states})} \nonumber\\
\text{Control Constraints}  & u_{c,min} \leq \boldsymbol{u}^{t}_{c,set,bas} \leq u_{c,max} \label{equ:cc-chiller-range-2},\\
    & \boldsymbol{C}^t_{reg} = \Psi (\boldsymbol{u}^{t}_{c,set,bas}) \label{equ:cc-fr-capacity},\\
    & \boldsymbol{u}^{t}_{c,set} = \boldsymbol{u}^{t}_{c,set,bas} + \widetilde{\boldsymbol{r}}^t\Delta \boldsymbol{u}^t_c \label{equ:cc-chiller-set-2},\\
    & \text{Eq.~(\ref{equ:cc-chiller-range}), Eq.~(\ref{equ:cc-storage-set}), Eq.~(\ref{equ:cc-storage-range})} \nonumber
\end{eqnarray}

Compared with the \textbf{Problem 1}, the cost function of the \textbf{Problem 2} has an additional term, the regulation revenue $R_{rev}$ from the regulation market. It is computed as follows:

\begin{equation}    
    R^{t}_{rev} = \widetilde{\boldsymbol{p}}^{t}_{rm}\boldsymbol{C}^{t}_{reg}=\sum^{t+h}_{i=t} \widetilde{p}^i_{rm}C^i_{reg}    
\end{equation}
where $\widetilde{p}_{rm}$ is the predicted price signal from the regulation market.

The constraint in Eq.~(\ref{equ:cc-chiller-range-2}) describes the physical allowable range of the design variable $\boldsymbol{u}^{t}_{c,set,bas}$.
Eq.~(\ref{equ:cc-fr-capacity}) defines the calculation of FR capacity from $\boldsymbol{u}^{t}_{c,set,bas}$ and the detailed description of the estimation method $\Psi$ is illustrated in Section~\ref{subsec:chap6-fr-bids}.
The real time chiller capacity setpoint trajectory $\boldsymbol{u}_{c,set}$ is calculated from Eq.~(\ref{equ:cc-chiller-set-2}) and its range is constrained by Eq.~(\ref{equ:cc-chiller-range}).
$\widetilde{\boldsymbol{r}}^t$ is the predicted regulation signal. 
Because the regulation signal is typically stochastic, and energy-neutral over a long period, a few research proposed to use historical signal for predictions, where it is claimed that if the FR resources can follow historical signal, they should be able to follow future signals~\cite{fabietti2018experimental,gorecki2017experimental}. 
In this paper, we use the historical signals as predictions.
$\Delta \boldsymbol{u}^t_c$ is the adjustable symmetric range of chiller capacity based on $\boldsymbol{u}^{t}_{c,set,bas}$, and illustrated in Eq.~(\ref{equ:chap6-duc}).
The storage charging rate setpoint is then computed and constrained by Eq.~(\ref{equ:cc-storage-set}) and Eq.~(\ref{equ:cc-storage-range}), respectively.

\subsubsection{Estimation of FR Capacity}\label{subsec:chap6-fr-bids}
The regulation capacity bid is determined hourly, considering the IT servers and cooling system.
For the server aggregator, the power consumption, as defined in Eq.~(\ref{equ:chap5-power-servers}), increases as the aggregated frequency $f_{agg}$ increases.
Therefore, the minimum and maximum server power can be estimated using Eq.~(\ref{equ:p_server_min}) and Eq.~(\ref{equ:p_server_max}).

\begin{align}
    P_{agg,min} = P_{agg}(f_{min}) \label{equ:p_server_min}, \\
    P_{agg,max} = P_{agg}(f_{max}) \label{equ:p_server_max}. 
\end{align}

The baseline server power consumption when the aggregator operates at a frequency of $f_{bas}$ can be calculated as 
\begin{equation}
    P_{agg,bas} = P_{agg}(f_{bas}).
\end{equation}

The upward and downward capacity of the aggregator then can be estimated as
\begin{align}
    C_{agg,reg,up} = \max(P_{agg,max} - P_{agg,bas}, 0),\\
    C_{agg,reg,do} = \max(P_{agg,bas} - P_{agg,min}, 0).
\end{align}

The symmetric regulation capacity is then obtained as 
\begin{equation}
    C_{agg,reg} = \min(C_{agg,reg,up},C_{agg,reg,do}).
\end{equation}

For a cooling system with a TESS, the regulation capacity is mainly related to the chiller capacity since the chiller power is significantly larger than the power of the cooling towers and condense water pumps.
Here a simple method is used to identify the available regulation range of chillers as shown in Eq.~(\ref{equ:chap6-chiller-fr-reg}) and the other cooling equipment is ignored.
$COP$ is the coefficient of performance of the chiller, and is related to temperatures and flow rates of both evaporators and condensers.
To simplify the implementation, the nominal COP is used in the formula.

\begin{equation}\label{equ:chap6-chiller-fr-reg}
    C_{c,reg} = min(\frac{u_{c,max}-u_{c,set,bas}}{COP},\frac{u_{c,set,bas}-u_{c,min}}{COP}).
\end{equation}

Therefore, the total regulation capacity bid can be estimated as 
\begin{equation}\label{eq:fr-cap-bid}
    C_{reg} = C_{agg,reg} + C_{c,reg}.
\end{equation}

\subsection{Stage 3: Tracking Controller}\label{subsec:stage3}
\begin{figure}[htbp]
    \centering
    \includegraphics[scale=0.5]{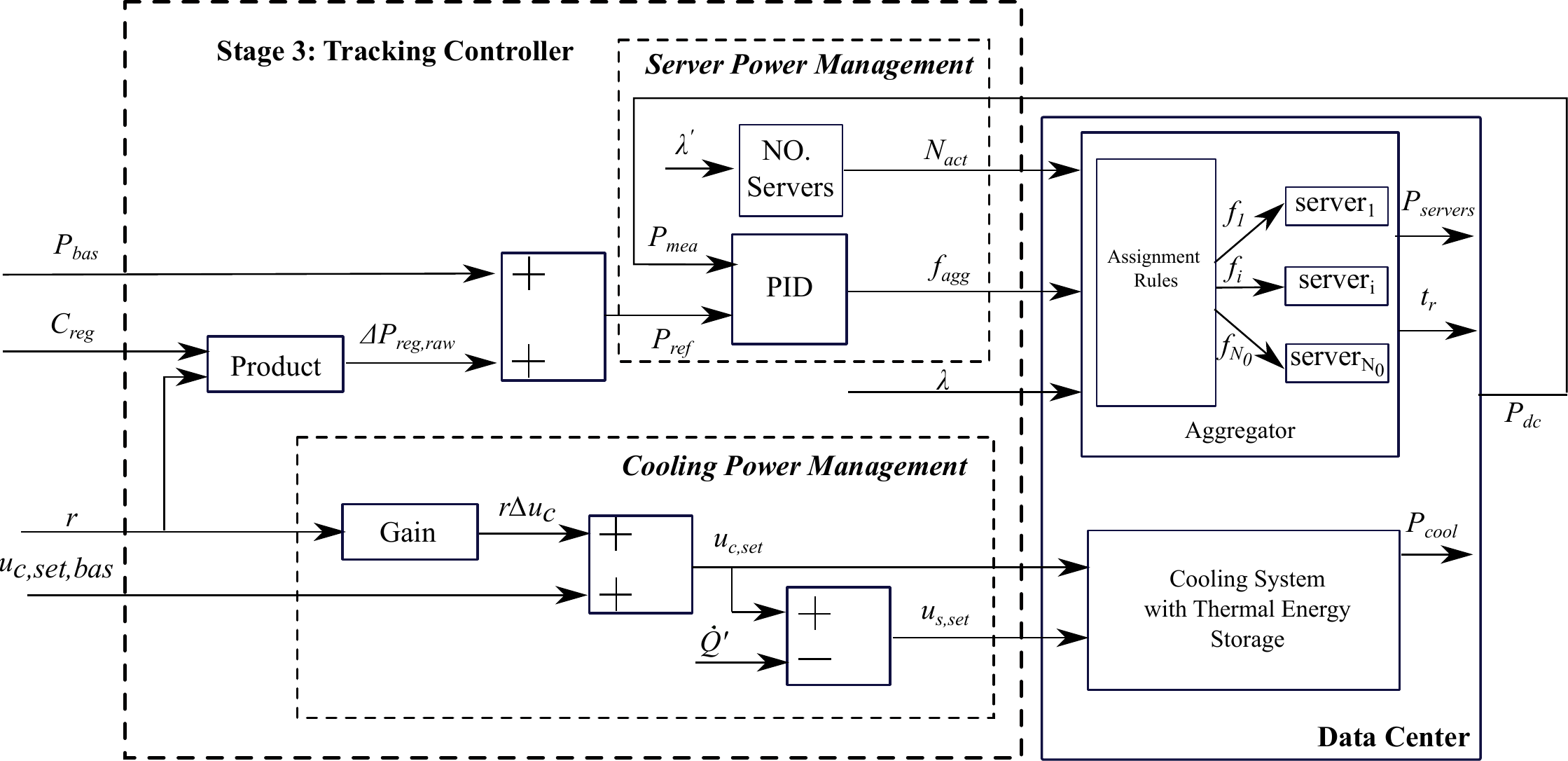}
    \caption{Synergistic control strategy for FR service in a data center with chilled-water TESS}
    \label{fig:chap6-frcontrol}
\end{figure}

The \textit{Stage 3} controller provides a control strategy to enable FR provision and the detailed design is shown in Figure~\ref{fig:chap6-frcontrol}.
This controller requires the following inputs: $P^t_{bas}$ from the stage 1 controller, $C^t_{reg}$ and $u^t_{c,set,bas}$ from the \textit{Stage 2} controller, and the regulation signal $r$ from electrical markets.
This strategy is composed of two major parts: 
The first one is \emph{Server Power Management}, where an aggregator is adopted to represent the aggregated performance of servers in the data center.
The clock frequency of the server aggregator can be directly changed by a Proportional-Integral-Derivative (PID) controller in order to follow the FR signal. 
Based on that, the desired frequencies for individual servers will be determined by a set of predefined assignment rules which will then be propagated to all servers.
The second one is the \emph{Cooling Power Management}, which adjusts the chiller capacity and storage charging rate to respond to the FR signal.

The \emph{Server Power Management} first determines the number of required active servers in the aggregator $N_{act}$ based on the predicted workload $\lambda^{'}$ in the next time step (e.g., one hour ahead). 
Then a closed-loop control using a PID controller is utilized to minimize the error between the measured total power usage $P_{dc}$ and the reference power $P_{ref}$ by adjusting the aggregated frequency of the server aggregator.
Meanwhile, the \emph{Cooling Power Management} applies an open-loop control to adjust the cooling system power usage in response to the received FR signal. This is completed by resetting the chiller capacity setpoint and the storage charging rate setpoint. 

The server aggregator receives the aggregated frequency $f_{agg}$ and the required number of active servers $N_{act}$ from the FR controller. 
Assuming there are $N_{0}$ number of servers in the data center, the server aggregator then calculates the CPU frequency $f_i$ for an individual server $i$ based on predefined assignment rules. 
The cooling system receives the chiller capacity setpoint from the FR controller.
Both the IT system and the cooling system respond in such a way that their total power $P_{dc}$ is adjusted to track the reference power $P_{ref}$.

For the aggregator, there are several assignment rules to control the individual server's frequency~\cite{wang2019frequency}.
We can also represent the aggregated server power $P_{agg}$ of all servers under an assignment rule using a simplified model~\cite{li2013data}. 
This paper has adopted and improved that approach. The details are presented in Section~\ref{subsec:chap6-server-management}.
For the FR controller, more details are described in the rest of this section.

\subsubsection{Server Power Management}\label{subsec:chap6-server-management}
The servers in the data center are represented by an aggregator, which is characterized by the total active servers $N_{act}$, and the aggregated frequency $f_{agg}$ as shown in Section~\ref{subsubsec:chap6-aggregator-model}. 
Based on these two parameters, the aggregator can output the total power of the servers $P_{servers}$ and the average service response time $t_r$.
The \emph{Server Power Management} is used to determine $N_{act}$ and $f_{agg}$ at each time step based on the normalized raw FR signal $r$ ranging from -1 to 1, and the incoming actual workload $\lambda$. 

\textbf{\textit{Number of Active Servers. }}
The following condition says that the service capability  $N_{act}\mu$ in the data center should be greater than the workload) $\lambda$ to ensure stability of the IT service:
\begin{equation}
    N_{act}\mu > \lambda, \label{equ:chap5-service-stability}
\end{equation}
where $\mu^t$ is the actual service rate, which denotes the number of requests that a single server can process every second.
The service rate is typically proportional to the server's CPU frequency, as defined in Eq.~(\ref{equ:chap6-service-rate})~\cite{wang2019frequency,li2013data}.

After introducing a scaling factor $\gamma$ and a FR flexibility factor $\beta$, the number of active servers can be determined as~\cite{fu2019assessments}:
\begin{equation} 
    N_{act} = \ceil{\beta\frac{\gamma\widetilde{\lambda}}{k}}, N_{act} \in [0,N_{0}]. \label{equ:chap5-Na-safe}
\end{equation}
where the operator $\ceil{x}$ is the ceiling function that gives the least integer greater or equal to $x$ and $\widetilde{\lambda}$ is the predicted workload. Here we use the mean workload of the current time step as the prediction. 
$\gamma$ describes the server design redundancy~\cite{li2013data}. 
The $\gamma$ is set to greater than 1. 
If $\gamma = 1$, it means all the CPU clock frequencies need to be set at the maximum level just to serve the average workload, which limits the potential of FR.
$\beta$ introduces FR flexibility to the system, and the greater $\beta$ is, the more servers are activated for a specific workload~\cite{fu2019assessments}.
$k$ is a constant parameter, denoting the aggregator nominal service rate.

\textbf{\textit{Aggregated Frequency. }}
The aggregated frequency $f_{agg}$ is regulated by a PID controller to track the reference power $P_{ref}$ that is calculated as
\begin{align}
    \Delta P_{reg,raw} = rC_{reg}, \\
    P_{ref} = P_{bas} + \Delta P_{reg,raw},
\end{align}
where $\Delta P_{reg,raw}$ is the raw regulation power signal.

The frequency $f_{agg}$ is then determined by the PID controller as follows.
\begin{equation}
    f_{agg} = K_{p}e(t) + K_{i}\int_{0}^{t} e(x)dx + K_{d}\dv{e(t)}{t}, ~f_{agg} \in [f_{min}, f_{max}],
\end{equation}
\begin{equation}
    e(t) = P_{ref} - P_{dc}.
\end{equation}

In the above equations, $K_p$, $K_i$, and $K_d$ denote the coefficients for the term P, I, and D respectively. 
$e$ is the control errors between $P_{ref}$ and $P_{dc}$.
The maximum aggregated frequency is 1, while the minimum aggregated frequency varies based on the number of active servers due to quality of service (QoS) restraints. 
One way to obtain the minimum aggregated frequency is to solve the following optimization problem.

\begin{equation}\label{equ:chap5-opt-fmin} 
\begin{aligned}
\min_{\rho} \quad & f = \frac{\widetilde{\lambda}}{kN_{act}\rho}, \\ 
s.t. \quad & 0 \leq \rho \leq 1, \\
           & t_{r} \leq t_{r,u}, 
\end{aligned}
\end{equation}

\noindent where $\rho$ is the utilization rate as defined in Eq.~(\ref{equ:chap6-utilization-rate}), $t_r$ is the service response time as calculated in Eq.~(\ref{equ:chap6-response-time}) and $t_{r,u}$ is the maximum response time allowed by the data center.

Utilizing the models in Section~\ref{subsubsec:chap6-aggregator-model} and rearranging Eq.~(\ref{equ:chap6-service-rate}) to Eq.~(\ref{equ:chap6-response-time}), we can get the response time $t_r$ as a function of the utilization rate $\rho$:

\begin{equation}\label{equ:chap5-tr-ur}
    t_r(\rho) = \frac{\rho}{\lambda}[N_{act}+\frac{C_A^2+C_B^2}{2(1-\rho)}Pr(\rho)],
\end{equation}

\noindent It is straightforward to show that 

\begin{equation}
    \dv{t_r(\rho)}{\rho} > 0.
\end{equation}

Thus, the above-mentioned optimization problem can be solved at each time step as:
\begin{equation}\label{equ:fmin}
    f_{min} = \frac{\lambda}{kN_{act}\rho^{*}},
\end{equation}
where $\rho^{*}$ is the optimal utilization rate, and $\rho^{*}$ should satisfy the nonlinear relationship shown as:

\begin{equation}
    t_r(\rho^{*})-t_{r,u}=0.
\end{equation}

\subsubsection{Cooling Power Management}\label{subsec:chap6-cooling-management}
The \emph{Cooling Power Management} modulates the cooling system response to the FR signal.
For a cooling system with a TESS, the chiller capacity responds to the FR signal from the electric market as shown in Eq.~(\ref{equ:chap6-ucset-fr}).
\begin{equation}\label{equ:chap6-ucset-fr}
    u_{c,set} = u_{c,set,bas} + r\Delta u_c,
\end{equation}
where $u_{c,set}$ is the chiller capacity updated on a 4-second basis, and $u_{c,set,bas}$ is a predefined base chiller capacity that can be scheduled to minimize the data center operational costs on an hourly basis. 
The $\Delta u_c$ is the regulation range of the chiller capacity, and it could be estimated from Eq.~(\ref{equ:chap6-duc}).

\begin{equation}\label{equ:chap6-duc}
    \Delta u_c = \min (u_{c,max} - u_{c,set,bas}, u_{c,set,bas}-u_{c,min}) 
\end{equation}

After obtaining $u_{c,set}$, the charging rate of the storage $u_{s,set}$ is then determined by Eq.~(\ref{equ:cc-storage-set}).
When $u_{c,set}$ is greater than the predicted cooling load $\widetilde{\dot{q}}$, then the storage charges at a rate of $u_{s,set}$.
When $u_{c,set}$ is less than $\widetilde{\dot{q}}$, then the storage discharges at a rate of $u_{s,set}$.

\section{Data Center Modeling}\label{sec:dc-model}
\subsection{Server Aggregator}\label{subsubsec:chap6-aggregator-model}
An aggregated server model described in~\cite{li2013data} is adopted here. This model can output the real-time power and service response times based on CPU frequency, workload arrival rate, and the number of active servers.

\begin{equation}\label{equ:chap5-power-servers}
P_{agg} = \lambda\sum_{0}^{2} b_if_{agg}^i + \sum_{0}^{1} c_jN_{act}^j,
\end{equation}

\noindent where $b_i$ and $c_j$ are constant coefficients that can be obtained from curve fitting techniques. 
Since most reported servers consume larger energy as the CPU frequency increases, we constraint $b_i$ and $c_j$ by:
\begin{align}\label{equ:chap5-bc}
b_i > 0, i >= 1 \\
c_j > 0, j > =1
\end{align}

The QoS of the data center is measured by the average response time. 
The workloads are modeled as GI/G/m queues, which assumes a general distribution with independent arrival time and a general distribution of service time~\cite{bolch2006queueing}. 
The total time that a job spends in the queuing system is known as the response time, which consists of service time $t_s$ and waiting time $t_w$
The average response time model is adopted from \cite{bolch2006queueing}. 
Details are shown as follows.

\begin{align}
    \mu = kf_{agg} \label{equ:chap6-service-rate}\\
    t_{s} = \frac{1}{\mu} \\
    \rho = \frac{\lambda}{N_{act}\mu}, 0 \leq \rho \leq 1 \label{equ:chap6-utilization-rate}
\end{align}
\begin{equation}
    Pr = \begin{cases}
	\frac{\rho^{N_{act}} + \rho}{2},& \rho \geq 0.7 \\
	\rho^{\frac{N_{act}+1}{2}}, & \rho < 0.7
	\end{cases}
\end{equation}
\begin{align}
    t_{w} = \frac{C_A^2+C_B^2}{2N_{act}}\frac{Pr}{\mu(1-\rho)} \\
    t_{r} = t_{s} + t_{w} \label{equ:chap6-response-time}
\end{align}
In the above equations, $\rho$ is the average utilization of the server, representing the fraction of occupied time, $Pr$ is approximated probability that an arriving job is queued, $C_A$ and $C_B$ are constant coefficients reflecting the type of data centers.

The data center servers generate the heat to the room and the heat flow as the cooling load is estimated using Eq.~(\ref{equ:cool-load}).
\begin{equation}\label{equ:cool-load}
    \dot q = P_{agg}
\end{equation}

\subsection{Stratified Chilled Water Tank}\label{subsubsec:chap6-tank-model}
\begin{figure}[htbp]
    \centering
    \includegraphics[scale=1]{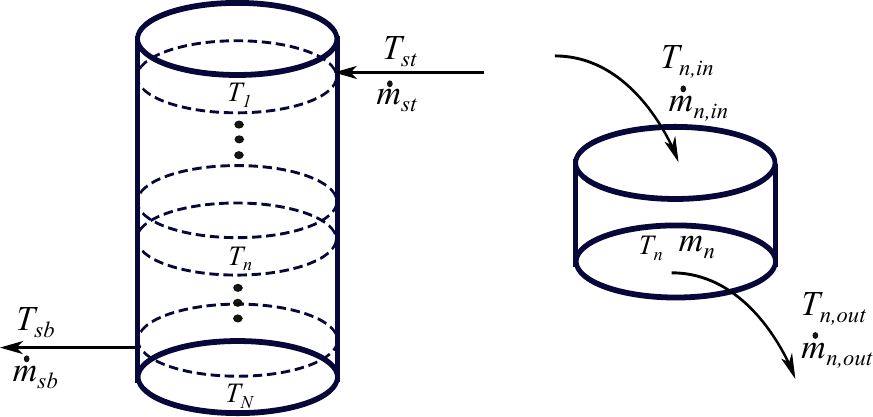}
    \caption{Finite volumes in stratified tank} 
    \label{fig:chap6-tank}
\end{figure}

The stratified tank is numerically modelled as a series of evenly-connected finite volumes as shown in Figure~\ref{fig:chap6-tank}~\cite{wischhusen2006enhanced}.
In volume $n$, the mass balance and energy balance equations for an incompressible medium with constant density (e.g., liquid water) are described in Eq.~(\ref{equ:chap6-tank-mass}) and Eq.~(\ref{equ:chap6-tank-energy}), where $m$ is the mass of the volume, $\dot{m}$ is the mass flowrate, $T$ is the temperature, $C_p$ is the specific heat of the fluid. 
The subscripts $n$, $in$, $out$ and $s$ represent the index, inlet, outlet, and source heat of the volumes respectively.
A discretisation method developed in~\cite{wischhusen2006enhanced} is used to solve Eq.~(\ref{equ:chap6-tank-energy}).

\begin{equation}\label{equ:chap6-tank-mass}
    \dv{m_n}{t} =  \dot{m}_{n,in}- \dot{m}_{n,out} = 0
\end{equation}

\begin{equation}\label{equ:chap6-tank-energy}
    m_{n}C_{p}\dv{T_{n}}{t} = \dot{m}_{n,in}C_{p}T_{n,in} - \dot{m}_{n,out}C_{p}T_{n,out} + \dot{q}_{n,s}
\end{equation}

The term ``SoC" is used to describe the level of charge of a chilled water stratified tank compared to its nominal capacity.
The SoC is 1 when the tank is fully charged and 0 when the tank is fully depleted.
SoC is difficult to measure, but it can be estimated from the tank temperatures. 
In this section, the average tank temperature $T_{s,avg}$ among different stratified volumes are used to estimate the current SoC as shown in Eq.~(\ref{equ:chap6-soc}).

\begin{equation}\label{equ:chap6-soc}
    SoC = \frac{T_{st,0}-T_{s,avg}}{T_{st,0}-T_{sb,0}},
\end{equation}
where $T_{st,0}$ and $T_{st,0}$ are the return temperature at the top of the tank and the supply temperature at the bottom of the tank under nominal conditions.

The charging rate $u_s$ during charging and discharging processes is measured as:
\begin{equation}
    u_s = \dot{m}_{sb}C_{p}(T_{st}-T_{sb}).
\end{equation}
The $\dot{m}_{sb}$ is the mass flowrate at the bottom of the tank.
While charging, the cold water flows into the tank and $\dot{m}_{sb}$ is set above zero.
While discharging, the cold water flows out of the tank and the $\dot{m}_{sb}$ is set below zero.
The $T_{st}$ is the temperature of the water flowing in or out of the top of the tank.
The $T_{sb}$ is the temperature of the water flowing in or out of the bottom of the tank.

\subsection{Other Models for Cooling Systems}
\subsubsection{Operating Mode Control}\label{subsubsec:chap6-operating-mode}
The chilled-water storage system for a data center can usually operate in four operating modes: $M_1$, charging storage while meeting loads; $M_2$, meeting loads from storage only; $M_3$, meeting loads from storage and chillers; $M_4$, meeting loads from chillers only.
The scheduled storage charging rate setpoint $u_{s,set}$ and chiller capacity setpoint $u_{c,set}$ are used to switch the cooling system among different operating modes.
The staging conditions for entering each mode are listed in Table~\ref{tab:chap6-transition-modes}.
Each staging condition has a delay time $\Delta t$ that prevents short cycling of cooling equipment and it is set to 5 minutes.

\begin{table}[htbp]
\centering
\caption{Staging conditions among different operating modes}
\label{tab:chap6-transition-modes}
\begin{tabular}{@{}cc@{}}
\toprule
Operating Modes     & Staging Conditions      \\ \midrule
$M_1$    & $u_{c,min} \leq u_{c,set} \leq u_{c,max}\text{ for } \Delta t$, \text{and } $0<u_{s,set}\leq u_{s,max} \text{ for } \Delta t$  \\
$M_2$ & $0 \leq u_{c,set} < u_{c,min} \text{ for } \Delta t$, \text{and }$u_{s,min} \leq u_{s,set} < 0 \text{ for } \Delta t$ \\
$M_3$ & $u_{c,min} \leq u_{c,set} \leq u_{c,max}\text{ for } \Delta t$, \text{and }$u_{s,min} \leq u_{s,set} < 0 \text{ for } \Delta t$ \\ 
$M_4$ & $u_{c,min} \leq u_{c,set} \leq u_{c,max}\text{ for } \Delta t$, \text{and }$u_{s,set} = 0 \text{ for } \Delta t$\\ \bottomrule
\end{tabular}%
\end{table}

The chilled water supply temperature is adjusted to track $u_{c,set}$.
When the cooling capacity is required to increase, the supply temperature is decreased, and vice versa. 
For the tank, the pressure-sustaining valves are modulated to regulate the water flow so that the charging rate $u_s$ can track their references $u_{s,set}$.
During the charging process, a PI controller is used to track the errors between the actual charging rate $u_s$ and the reference charging rate $u_{s,set}$ by adjusting the valve position of the PSV\textsubscript{2}.
The changes of the valve position lead to the changes of water flowing into the tank, which eventually leads to changes in the actual charging rate.
The same control loop is also used for the discharging process, but with modulating the PSV\textsubscript{1} instead.

\subsubsection{Other Models}
Other components in the cooling system, including air handler unit, chiller, cooling tower etc., are modeled in a previously developed Modelica-based environment~\cite{fu2019equation-b,fu2019modelica, fu2018modelica,fu2019equation}. 
The models have been well demonstrated for building energy system control.
The prediction errors of the models such as air handler unit, chiller, cooling tower, and pumps are within 6\% compared with measurement data.

\section{Case Study}\label{sec:chap6-case-study}
This case study evaluates the performance of the proposed multi-market optimization framework through numerical experiments.
The studied system is schematically shown in Figure~\ref{fig:chap6-charging-discharging} and modeled in the Modelica environment.
Note that in this case study, the testbed of the cooling system and the MPC controller use the same set of models as described in Section~\ref{sec:dc-model}, which assumes perfect predictions in the MPC controllers.
The evaluation is performed under the real-time energy market and regulation market in PJM. 
Only dynamic regulation (e.g., RegD) is studied for FR service in the regulation market because its price is usually significantly higher than traditional regulation.

This case study considers three different scenarios:
\begin{itemize}

    \item \textit{BL (BaseLine)}: the baseline system utilizes a storage-priority control strategy as described in~\cite{henze1997development} to enable the tank for load shifting. 
    The tank is charged during the off-peak period and discharged during the on-peak period.
    Here we set the on-peak period as hours from 11am to 7pm. After simulation, we found that the baseline control results in a power demand of 2,148 kW.

    \item \textit{BL+MM (Multi-Market)}: this scenario allows FR service in the \textit{BL} system.
    Instead of using the \textit{Stage 1} controller to purchase power baseline, this scenario uses the power profile predicted under the control of the storage-priority strategy as the base power for FR service.
    The regulation reserve schedule is estimated using the \textit{Stage 2} controller, and the real-time FR signal tracking is enabled by the \textit{Stage 3} controller. The demand limit in \textit{Stage 2} controller is set to same demand limit as \textit{BL}, that is, 2148 kW.

    \item \textit{OPBL (OPtimal BaseLine)+MM}: this scenario applies to the studied data center the proposed multi-market optimization framework in Section~\ref{sec:multi-market-control}.
    This scenario can minimize the operational costs from both energy market and regulation market. 
    The power demand in this scenario is set to 1990 kW, a lower value than that in \textit{BL}.
    
\end{itemize}

\subsection{Case Description}\label{subsec:chap6-case-description}
\begin{figure}[htbp]
    \centering
    \includegraphics[scale=0.4]{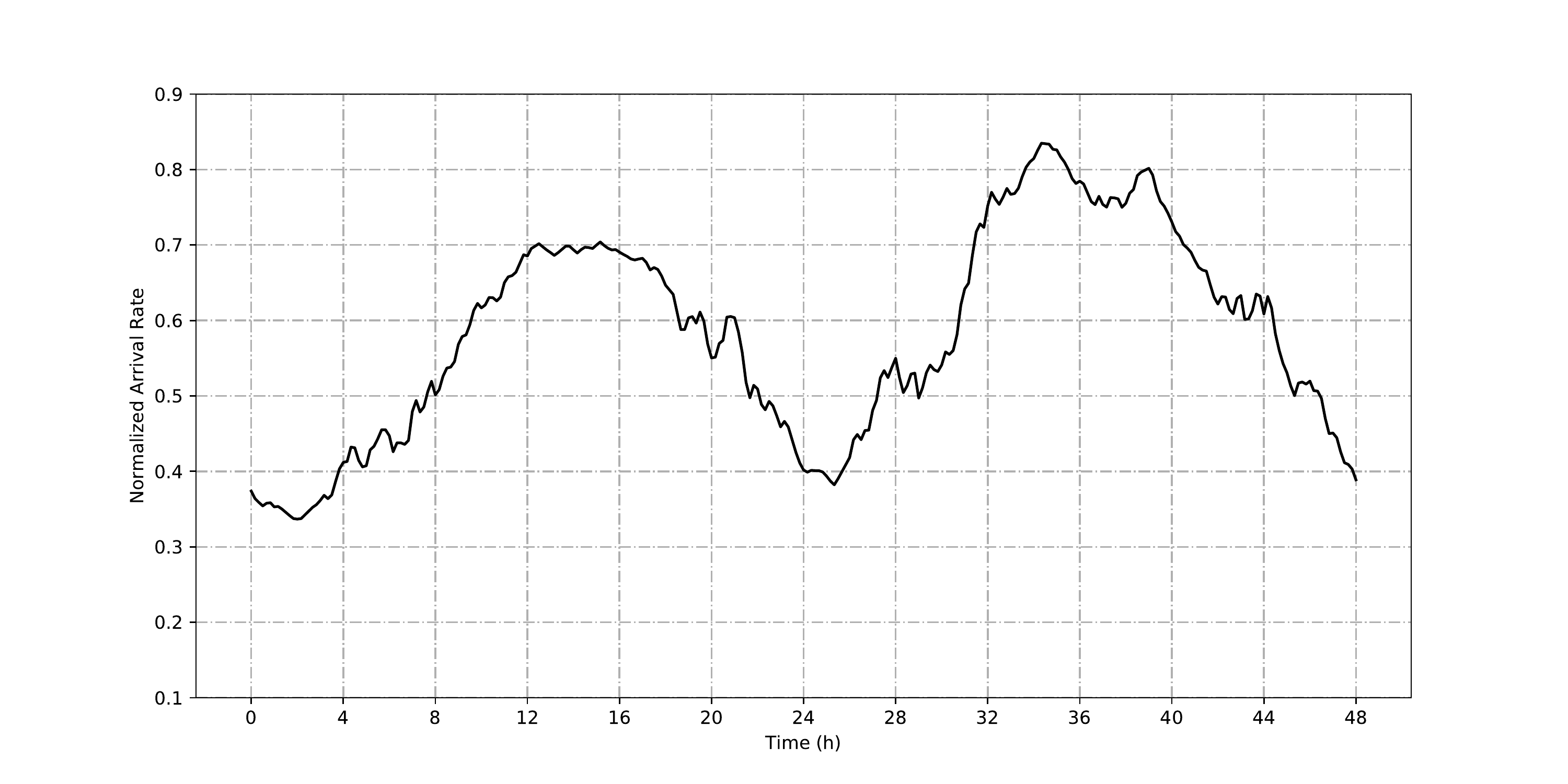}
    \caption{Two-day historical arrival rates in the data center}
    \label{fig:chap6-arrival-rate-multi-market}
\end{figure}

The schematic drawing of the studied system is shown in Figure~\ref{fig:chap6-charging-discharging} and the total nominal electrical load is 2,680 kW. 
For the IT system, the design number of servers is 16,000 with an individual nominal power of around 124 W. 
The design factor $\gamma$ is set to 1.5 \cite{li2013data}.
The calibrated coefficients for the server aggregator describing by the Eq.~(\ref{equ:chap5-power-servers}) are $b_{0}=0.016$, $b_{1}=1.60$, $b_{2}=0.14$, $c_{0}=0.01$ and $c_{1}=120.92$ using the method mentioned in~\cite{li2013data}.
A reported two-day request trace obtained from the Wikipedia webpage is normalized and used in this study (Figure~\ref{fig:chap6-arrival-rate-multi-market})~\cite{urdaneta2009wikipedia}.

There is only one chiller and one storage tank for cooling.
The nominal chiller capacity $C_{c}$ is 1,982 kW with a design COP of 5.8 and the thermal time constant of the chiller is set to 5 minutes.
The tank is sized to fully address a 4-hour nominal cooling load of 1,982 kW in the data center using Eq.~(\ref{equ:chap6-storage-capacity}). 
The minimum and maximum SoC of the tank are set to 0.05 and 0.95 respectively.
The chilled water supply temperature is regulated within $8.5 \pm 3.5$ \textdegree C.
The data center room temperature is bounded within $25 \pm 3$ \textdegree C.
The control input $u_{c,set,bas}$ and $u_{c,set}$ are constrained within $[0.05*C_{c},C_{c}]$ for simplification.

\begin{equation}\label{equ:chap6-storage-capacity}
    C_{s} = C_{c}*4*3600
\end{equation}

\begin{table}[htbp]
\centering
\small
\caption{Cooling system simulation parameters}
\label{tab:cooling-setup}
\resizebox{\textwidth}{!}{%
\begin{tabular}{@{}lllll@{}}
\toprule
Equipment            & Qty. & Parameter                   & Unit & Value \\ \midrule
Room                 & 1    & Cooling Load                & kW   & 1982  \\
Servers              & 16000 & Power                       & W    & 124   \\
Chiller              & 1    & Cooling Capacity            & kW   & 1982  \\
                     &      & Design COP                  &      & 5.8   \\
Chilled Water Tank   & 1    & Capacity                    & MJ   & 28550 \\
Primary Pump         & 1    & Head                        & kPa  & 120   \\
                     &      & Power                       & kW   & 7     \\
                     &      & Flowrate                    & kg/s & 52.6  \\
Secondary Pump       & 1    & Head                        & kPa  & 400   \\
                     &      & Power                       & kW   & 22.5  \\
                     &      & Flowrate                    & kg/s & 52.6  \\
Condenser Water Pump & 1    & Head                        & kPa  & 300   \\
                     &      & Power                       & kW   & 30    \\
                     &      & Flowrate                    & kg/s & 92.5  \\
Transfer Pump        & 1    & Head                        & kPa  & 66    \\
                     &      & Power                       & kW   & 4     \\
                     &      & Flowrate                    & kg/s & 52.6  \\
Cooling Coil         & 1    & Cooling Capacity            & kW   & 1980  \\
                     &      & Air Flowrate                & kg/s & 218.5 \\
                     &      & Water Flowrate              & kg/s & 52.6  \\
Supply Air Fan       &      & Head                        & Pa   & 622   \\
                     &      & Power                       & kW   & 210   \\
Cooling Tower        & 1    & Nominal Capacity            & kW   & 2320  \\
                     &      & Design Approach Temperature & K    & 4.4   \\
                     &      & Power                       & kW   & 86    \\ \bottomrule
\end{tabular}%
}
\end{table}

When providing FR service, the regulation flexibility factor $\beta$ is set to 1.1 based on a previous parametric study~\cite{fu2019assessments}.
The QoS the average response time of the data center service to guarantee the QoS is set to 6 ms.
The prediction horizon in the \textit{Stage 1} and \textit{Stage 2} controller is set to 12 hours, and the control horizon is set to 1 hour.
The demand limit is set to 1,990 kW, a lower value than the baseline demand 2,148 kW in $BL$.

\begin{figure}[htbp]
\centering
\includegraphics[scale=0.4]{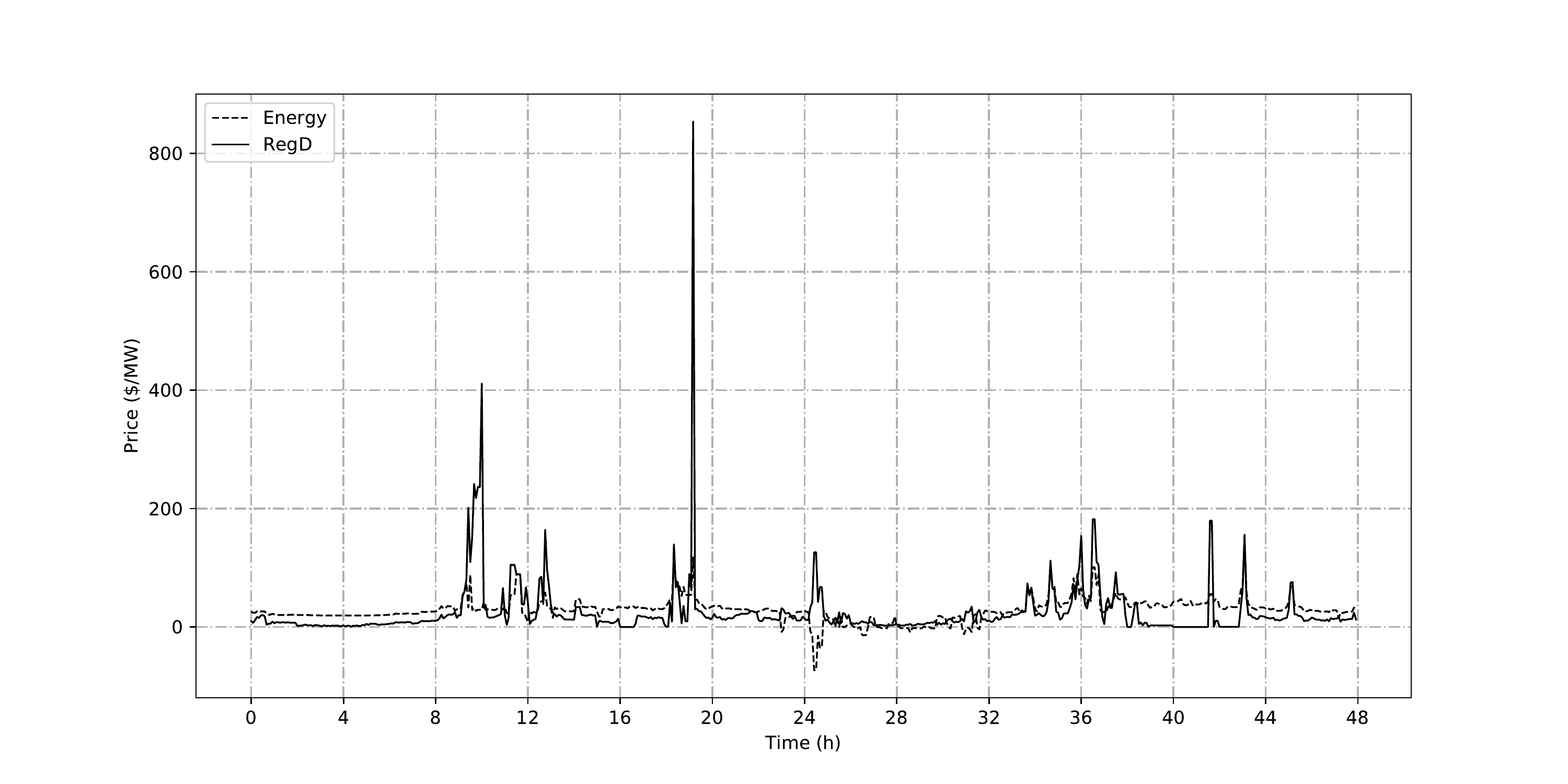}
\caption{Two-day historical real-time prices of PJM markets} 
\label{fig:chap6-prices}
\end{figure}

Figure~\ref{fig:chap6-prices} shows the two-day historical prices (7/1/2018 and 7/2/2018) from the energy market and regulation market in PJM.
The demand charge rate is set to 7.48 \$/kW. 
The historical RegD signal from the PJM market in the same two-days can be referred to Ref.~\cite{PJM2019selftest}.

\subsection{Results and Discussions}\label{subsec:chap6-results}
This section presents the simulation results, including the control performance of the proposed synergistic control strategy for FR provision in Section~\ref{subsubsec:chap6-fr-control}, and the results from the proposed multi-market optimization framework in Section~\ref{subsubsec:chap6-multi-market-optimization}.

\subsubsection{Frequency Regulation Control}\label{subsubsec:chap6-fr-control}
This section investigates the control performance of the proposed FR control strategy as defined in Section~\ref{subsec:stage3}, including the FR signal tracking performance, hourly control inputs, and hourly FR capacity bids. The \textit{OPBL+MM} is used as an example. 

\begin{figure}[htbp]
\centering
\subfigure[Hourly performance score]{%
\resizebox*{12cm}{!}{\includegraphics{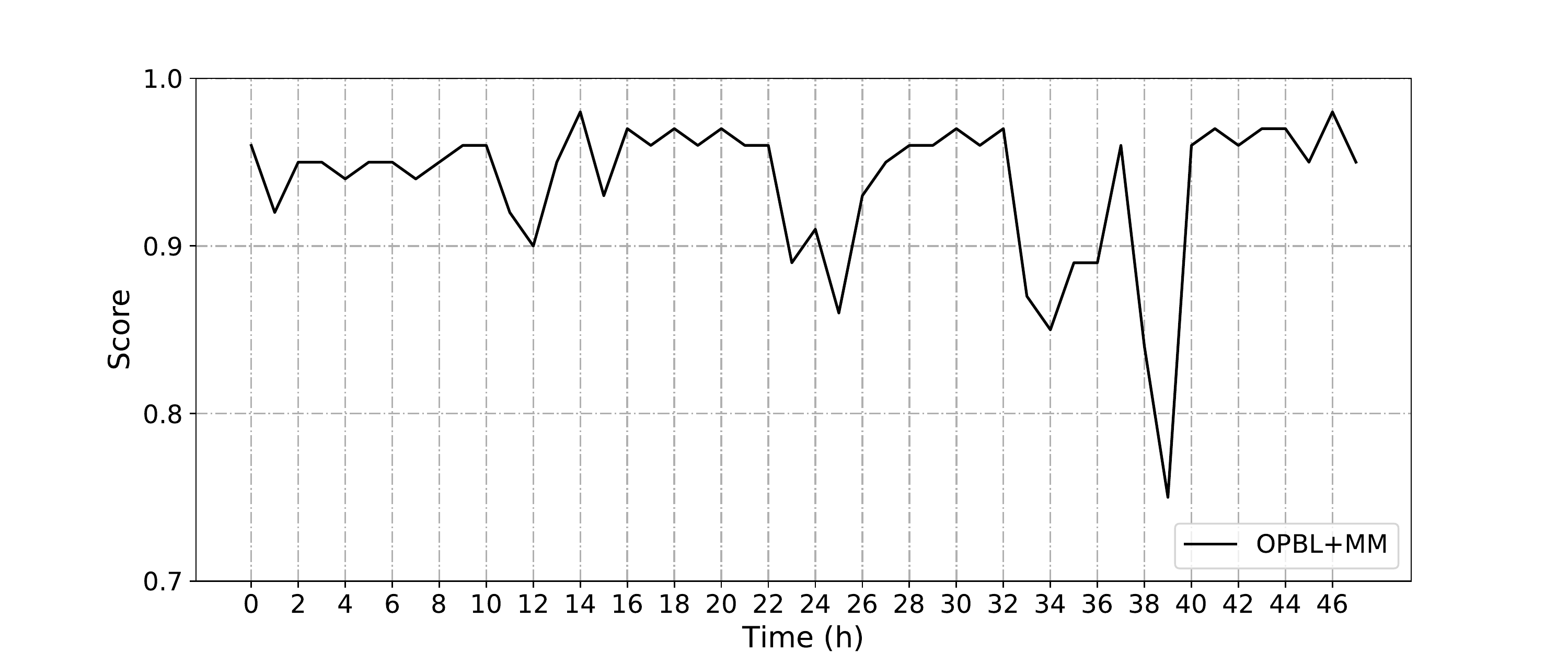}}}
\subfigure[Detail signal tracking]{%
\resizebox*{12cm}{!}{\includegraphics{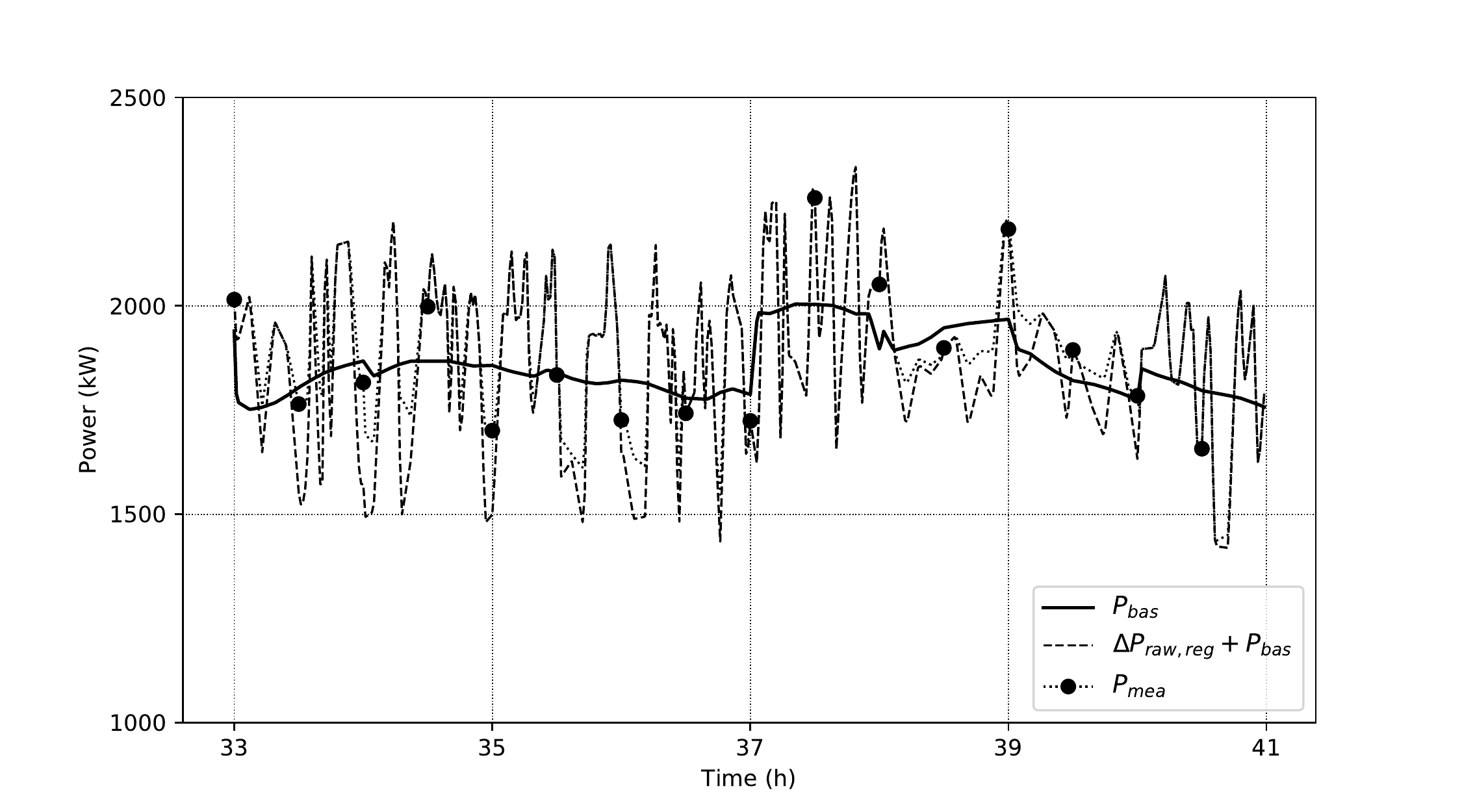}}}
\caption{FR signal tracking performance} 
\label{fig:chap6-fr-control-performance-bllsmm}
\end{figure}

During the simulated two days, the minimum hourly regulation performance score is 0.75, the maximum is 0.98, and the hourly average is 0.94 as shown in Figure~\ref{fig:chap6-fr-control-performance-bllsmm}.
The performance score is influenced by the regulation capacity bid and the system dynamics such as thermal dynamics in the cooling system and control delays such as in Section~\ref{subsubsec:chap6-operating-mode}.
PJM requires an initial score of 0.75 to enter the market and maintain 0.4 while participating. Thus, the proposed synergistic control strategy is qualified based on PJM's criteria.
At around hour 24 and hour 39, the performance score is lower than other periods. The primary reason is the overestimation of the regulation capacity bid, which leads to insufficient upward and downward regulation. 
Eq.~(\ref{eq:fr-cap-bid}) in the \textit{Stage 2} controller uses rules to estimate the regulation range, which eventually leads to underestimation or overestimation. 
If underestimated, the required regulation range is within the system's capability. Therefore, the system can generally provide good FR service. 
If the regulation capacity is overestimated, the system is asked to regulate its power within a range that it cannot physically provide, which leads to deteriorated FR service.
For example, at hour 39, the regulation capacity is so overestimated that the system cannot provide enough downward regulation. 
The servers work at their minimum frequency but still cannot reach the required lower limit.

\begin{figure}[htbp]
    \centering
    \includegraphics[scale=0.6]{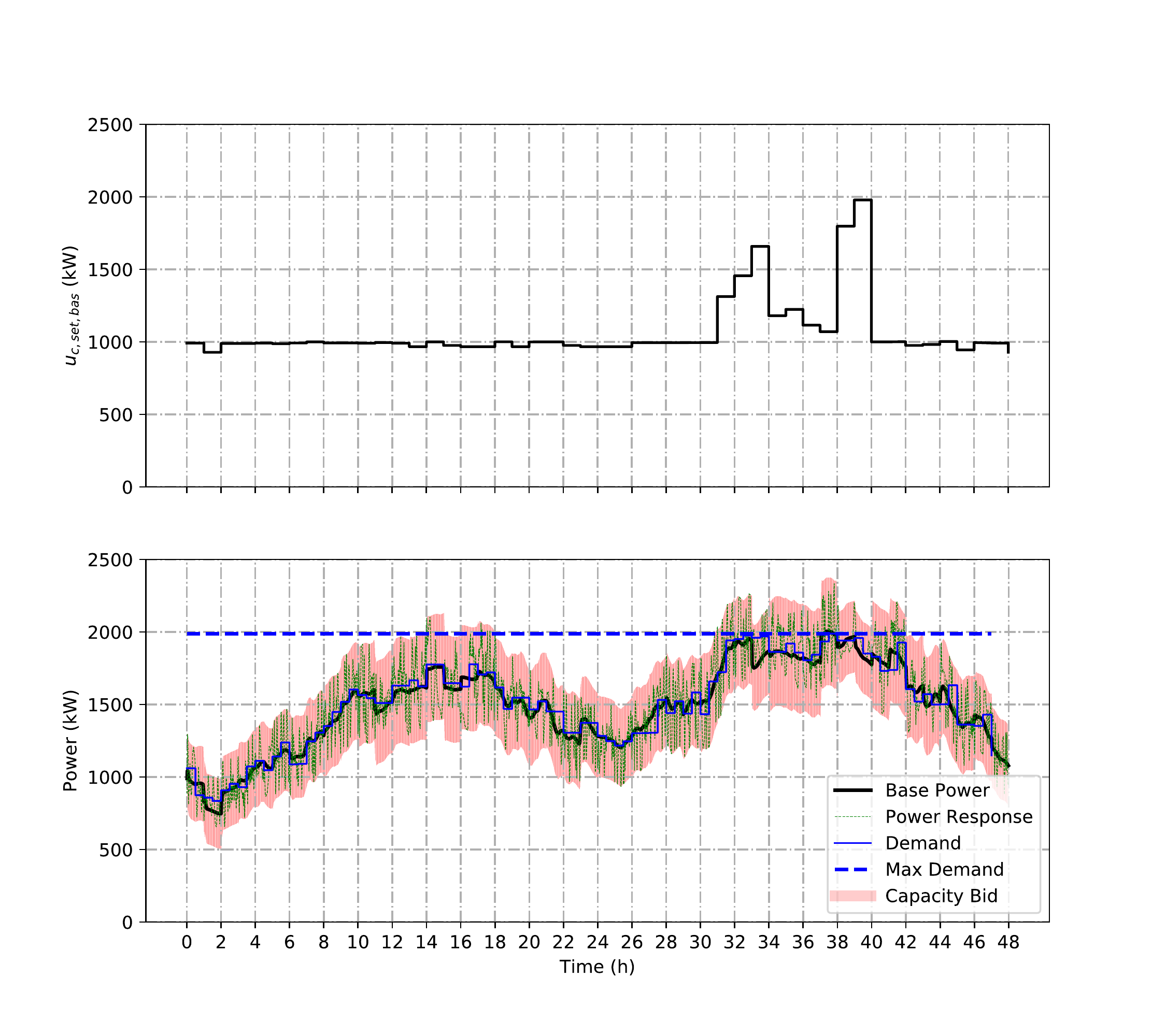}
    \caption{FR capacity in \textit{OPBL+MM}}
    \label{fig:chap6-fr-control-capacity-bloplsmm}
\end{figure}

Figure~\ref{fig:chap6-fr-control-capacity-bloplsmm} illustrates the detailed control signals and performance at each hour during the simulated two days.
The top figure shows the hourly control signal of the base chiller capacity setpoint $u_{c,set,bas}$.
The control signal is about 1,000 kW, half of the nominal capacity of the chiller, when the system power is far from its demand limit, e.g., hour 0 to hour 30.
When the system approaches its peak power due to peak workload, the control signal is significantly increased.
This happens because the small operating load leads to a power demand much less than the limit so setting $u_{c,set,bas}$ around 1,000 kW can provide the maximum FR capacity based on Eq.~(\ref{equ:chap6-chiller-fr-reg}) without violating the demand limit. Thus, maximum FR revenues are gained.
When the system reaches its demand limit, the FR capacity bid should be as small as possible by increasing the controlled chiller capacity. This avoids violating the demand limit as shown in the bottom figure.
For example, the FR capacity bid around hour 37 is so well controlled that the demand in that hour is at the limit.

A data center with a TESS can provide significant FR capacity.
The capacity bids can vary from a minimum of 242 kW (9\% of system nominal power) at hour 2 to a maximum of 378 kW (14\% of system nominal power) at hour 34 as shown in Figure~\ref{fig:chap6-fr-control-capacity-bloplsmm}.
The reason is that with storage installed, the system can bid its maximum FR capacity into the market most time, while the system without storage can only bid a capacity between 0 and the maximum to balance the FR revenues and the potential increase of energy and, especially, demand costs~\cite{fu2020multi}. 

\subsubsection{Multi-market Scheduling}\label{subsubsec:chap6-multi-market-optimization}

This part discusses the performance of the proposed multi-market scheduling framework.
Table~\ref{tab:chap6-opt-results-w-tes} shows the operational costs of two days in different scenarios.
The energy consumption in all scenarios are similar, but the total operational costs are different with up to 8.8\% cost savings by \textit{OPBL+MM}.

\begin{table}[htbp]
\centering
\caption{Comparison of two-day operational costs}
\label{tab:chap6-opt-results-w-tes}
\begin{tabular}{@{}cccccc@{}}
\toprule
Items & \textit{BL} & \textit{BL+MM} & \textit{OPBL+MM} \\ \midrule
Energy (MWh) &72.5 & 72.9 & 72.0 \\
Energy Cost (\$) & 2,212.3 & 2,197.5 & 2,173.6 \\
Demand (kW) & 2,148 & 2,148 & 1,990 \\
Demand Cost (\$) & 16,064.6 &16,064.6 & 14,885.2 \\
FR Cost (\$) & 0 & -385.4 & -388.3 \\
Total Cost (\$) & 18,276.9 & 17,876.7 & 16,670.5 \\ 
Relative Savings & 0\% & 2.2\% & 8.8\% \\ \bottomrule
\end{tabular}
\end{table}

The system with a TESS can obtain a significant amount of revenues by participating in energy regulation markets without compromising energy and demand costs.
For example, \textit{BL+MM} can harness \$385.4 in the regulation market, but with similar energy costs and demand costs as in \textit{BL}.
The slight differences in the energy and demand costs are caused by the FR signal profiles.  
Because the sum of the RegD signal over a long time period (e.g. 1 hour) is almost 0, providing FR service in the \textit{BL+MM} leads to a similar energy use and thus similar energy cost compared with the \textit{BL} where no FR service is provided.
By utilizing the demand cost defined in Eq.~(\ref{equ:chap6-demand-cost}), the data center can provide FR service without increasing monthly demand. Therefore no extra demand charges would be added to the utility bills.

\begin{figure}[htbp]
    \centering
    \includegraphics[scale=0.6]{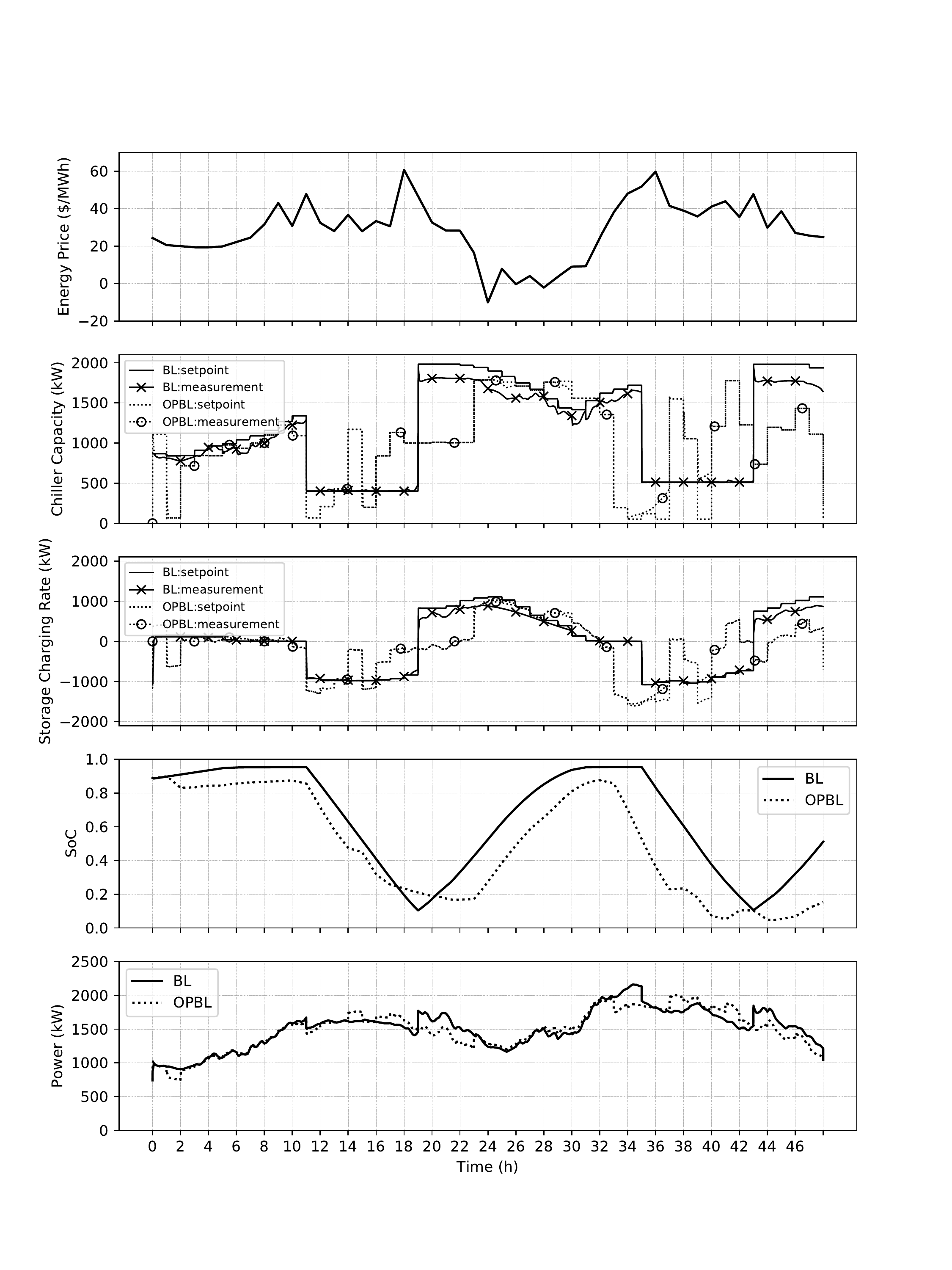}
    \caption{Control comparison of \textit{BL} and \textit{OPBL}} 
    \label{fig:chap6-blopls-blls}
\end{figure}

The \textit{OPBL+MM} can save \$1,186.2 (6.5\%) of total costs, compared with \textit{BL+MM}.
The savings are mainly from the different power baseline schedules.
Here we denote the purchased power baseline using the \textit{Stage 1} controller as \textit{OPBL}.
Figure~\ref{fig:chap6-blopls-blls} compares the control signals of the two baseline profiles \textit{OPBL} and \textit{BL}.
Using the conventional control strategy, \textit{BL} starts to charge the tank at hour 19 by setting the chiller capacity to its nominal value, the upper limit of the control signal. 
Note that the chiller can only provide a capacity of 1,800 kW at that hour, 10\% less than its setpoint, because the available chiller capacity varies when the chilled water supply temperature changes. 
At hour 19, the chilled water temperature is only 5 \textdegree C, which leads to an available capacity less than the nominal value.
However, \textit{OPBL} starts to charge the tank at hour 23 to avoid the high energy prices during hour 18 to hour 23.
The tank is fully charged to its maximum SoC (i.e., 0.95) in \textit{BL}, but it is only charged to about 0.87 in \textit{OPBL}.
This happens because the data center in these two days operates at its part load, so the tank does not need to be fully charged to meet the loads during the next on-peak.
Without charging the tank to its maximum level, about 0.9 MWh of energy can be saved in \textit{OPBL} in two days.

The different chiller capacity schedules result in different power profiles shown in the bottom figure. 
In \textit{BL}, since the storage is prioritized to shave peak power during 11am - 7pm, the maximum power demand of 2,148 kW happens at 10am - 11am during the second day.
With demand limiting strategy and predictive control in \textit{OPBL}, the peak demand happens at 8am-9am and decreases to 1,990 kW, which contributes the majority of operational cost reductions.

\section{Concluding Remarks}\label{sec:chap6-conclusions}
This paper proposed a new control strategy that enables FR service using a TESS in data centers and developed a multi-market optimization framework to minimize the operational costs.
Simulation results show that utilizing the TESS can not only reduce energy costs and demand charges, but also harvest FR revenue.  
Using the proposed multi-market optimization framework, a data center with a TESS during two days can save operational costs up to 8.8\% (\$1,606.4) compared to the baseline cost reduction of 0.2\% (\$38.7), 6.5\% (\$1,179.4) from demand reduction, and 2.1\% (\$338.3) from regulation revenues.

\section{Acknowledgments}
This research is financially funded by U.S. Department of Energy under the award NO. DE-EE0007688.
This work was also supported by the Assistant Secretary for Energy Efficiency and Renewable Energy, the U.S. Department of Energy under Contract No. DE-AC02-05CH11231.

This work emerged from IBPSA Project 1, an international project conducted under the umbrella of the International Building Performance Simulation Association (IBPSA). 
Project 1 will develop and demonstrate a BIM/GIS and Modelica Framework for building and community energy system design and operation.

\bibliography{manuscript}

\end{document}